\documentclass[fleqn,usenatbib]{mnras}

\usepackage{newtxtext,newtxmath}

\usepackage[T1]{fontenc}
\usepackage{ae,aecompl}
\usepackage{graphicx}
\usepackage{amssymb}
\usepackage{amsmath}
\usepackage{pifont}
\usepackage{times}
\usepackage{xcolor}

\usepackage{graphicx}	
\usepackage{amsmath}	
\usepackage{amssymb}	

\title[Characterizing Tidal Tails in Merging Galaxies]{Long Tidal Tails in Merging Galaxies and Their Implications}

\author[J.~Ren et al.]{
Jian Ren,$^{1,2}$\thanks{E-mail: renjian@pmo.ac.cn (JR); xzzheng@pmo.ac.cn (XZZ)}
X.~Z.~Zheng,$^{1,2 \color{blue}\star}$
David~Valls-Gabaud,$^{3}$
Pierre-Alain Duc,$^{4}$
Eric~F.~Bell,$^{5}$
\newauthor
Zhizheng~Pan,$^{1,2}$
Jianbo~Qin,$^{1,2}$
D.~D.~Shi,$^{1,2}$
Man~Qiao,$^{1,2}$
Yongqiang~He,$^{1,2}$
Run~Wen$^{1,2}$
\\
$^{1}$Purple Mountain Observatory, Chinese Academy of Sciences, 10 Yuanhua Road, Nanjing 210023, China\\
$^{2}$School of Astronomy and Space Sciences, University of Science and Technology of China, Hefei 230026, China\\
$^{3}$LERMA, CNRS, PSL, Observatoire de Paris, 61 Avenue de l'Observatoire, 75014 Paris, France\\
$^{4}$Observatoire Astronomique de Strasbourg, Universit\'e de Strasbourg, CNRS, 11 Rue de l'Universit\'e, F-67000 Strasbourg, France\\
$^{5}$Department of Astronomy, University of Michigan, 1085 South University Avenue, Ann Arbor, MI 48109, USA
}

\date{Accepted XXX. Received YYY; in original form ZZZ}

\pubyear{2020}

\begin{document}
\label{firstpage}
\pagerange{\pageref{firstpage}--\pageref{lastpage}}
\maketitle

\begin{abstract} 
We investigate the properties of long tidal tails using the largest to date sample of 461 merging galaxies with   
$\log(M_\ast/\rm M_\odot)\geq9.5$  within 0.2\,$\leq\,z\,\leq 1$ from the COSMOS survey in combination with {\it Hubble Space Telescope} imaging data.  Long tidal tails can be briefly divided into  three shape types: straight (41\,per\,cent), curved (47\,per\,cent) and plume (12\,per\,cent).  Their host galaxies are mostly at late stages of merging, although 31\,per\,cent are galaxy pairs with projected separations $d>20$\,kpc.  The high formation rate of straight tidal tails needs to be understood as the projection of curved tidal tails accounts for only a small fraction of the straight tails.   
We identify 165 tidal dwarf galaxies (TDGs),  yielding a TDG production rate of 0.36 per merger. Combined with a galaxy merger fraction and a TDG survival rate from the literature, we estimate that $\sim$\,5\,per\,cent of local dwarf galaxies are of tidal origin, suggesting the tidal formation is not an important formation channel for the dwarf galaxies. More than half of TDGs are located at the tip of their host tails. These TDGs  have stellar masses in the range of  $7.5\leq\log (M_\ast/\rm M_\odot)\leq9.5$ and appear compact with half-light radii following the $M_\ast$ - $R_{\rm e}$ relation of low-mass elliptical galaxies. However, their surface brightness profiles are generally flatter than those of local disc galaxies. Only 10 out of 165 TDGs have effective radii larger than 1.5 kpc and would qualify as unusually bright ultra-diffuse galaxies. 
\end{abstract}




\section{Introduction}

 In the $\Lambda$ Cold Dark Matter ($\Lambda$CDM) cosmological model,  cosmic structure and dark matter haloes grow in a hierarchical manner, placing galaxy mergers as major drivers of  galaxy formation and evolution \citep[e.g.][]{Hopkins2006}.  The mergers of galaxies not only drive the ex situ mass growth to form more massive galaxies \citep[e.g.][]{Bundy2009,Fakhouri2010, Oser2010, Moustakas2013, Rodriguez-Gomez2016,Martin2017}, but also  invoke gravitational torques/disruption to shape the  kinematic and physical properties of the merger remnants \citep[e.g.][]{Hopkins2007,Ellison2013,DePropris2014,Thorp2019}.  The past two decades have witnessed significant progress in our understanding of galaxy merger rate across cosmic age both observationally \citep{Patton2002, Conselice2003b, Lin2004, Lotz2008a, Jogee2009, Robaina2009, Lotz2011, Xu2012, Man2016, Wen2016, Mantha2018} and theoretically \citep{Stewart2009, Hopkins2010b, Fakhouri2010, Rodriguez-Gomez2015}.  Recently, more efforts have turned to characterizing galaxy mergers of different types and their essential roles in regulating the evolution of galaxies over a wide range of properties. 
 
Galaxy mergers are often classified into different types in terms of the properties of the merging galaxies, such as their mass ratio (major versus minor; \citealt{Bournaud2005,Lotz2010,Lotz2011,Ownsworth2014}) and gas fraction (wet versus dry; \citealt{Bell2006,Lin2008}). Major mergers, often defined by a mass ratio of $m/M>0.3$, have been widely studied in both simulations and observations. As the most violent process of  galaxy interactions, major mergers trigger intense star formation \citep[][]{Sanders1996,Mihos1996,Teyssier2010}, in at least some cases ignite active galactic nuclei (AGNs) \citep{Sanders1988,Bahcall1997,Treister2012,Barrows2017,Ricci2017} and induce morphological transformation from discs into spheroids \citep[e.g.][]{Barnes1992,Genzel2001,Naab2003,Lotz2008b,Bournaud2011}.

In contrast, minor mergers occur more often than major mergers and supply stars and gas into the primary galaxies.  Observational evidence suggests that minor mergers are dominantly responsible for the size growth of massive galaxies \citep[e.g.][]{Newman2012,Bluck2012,Hilz2013} and for triggering  star formation in at least the nearby universe \citep{Kaviraj2014}.  The minor mergers might even induce morphological transformation through disc instabilities for $z>1$ galaxies \citep[e.g.][]{Welker2014}.

Gas-rich mergers behave very differently from gas-poor mergers. At least some mergers drive gas inflows which fuel star formation and can trigger AGN activity. These gas inflows are expected to be important in influencing the growth of bulges \citep{Hopkins2009,Hopkins2010a}, but the relationship between merger parameters and bulge growth is far from straightforward \citep{Bell2017,Gargiulo2019}.  Disc survival and regrowth is expected to be impacted by gas content also, where even gas-rich major mergers are predicted to lead to disc-dominated remnants \citep{Robertson2006}. 

On the contrary, dry mergers of gas-poor galaxies proceed with dissipationless processes and sustain little star formation, being a key driver of the mass growth and structural evolution of  massive elliptical galaxies \citep{vanDokkum2005,Bell2006} and the Brightest Cluster Galaxies (BCGs) at $z<1$ \citep[e.g.][]{Lin2013}. Moreover, the internal structure and orbital parameters (prograde versus retrograde) of the merging galaxies also influence the properties of the post-merger galaxies  \citep{Darg2010,Martin2018}.  A comprehensive picture of shaping the detailed properties of present-day galaxies is still lacking.

 Indeed, tidal features generated in a galaxy merger strongly depend on the mass ratio, gas fraction, internal structure and orbital parameters of the merging galaxies, and thus can be used to characterize the merger \citep{Duc2013,Barnes2016}. For instance,  bridges and long tidal tails can be generated in major mergers between equal-mass discs \citep{Toomre1972,Barnes1992}, and tidal streams are often produced in the minor merger events such that satellite galaxies fall into a more massive galaxy \citep{Martinez-Delgado2010}.  Such tidal features are frequently detected around massive galaxies, tracing the merger events happened in the recent past \citep{Duc2015,Morales2018,Hood2018,Kado-Fong2018}.  In particular, long tidal tails are exclusively created in the major mergers involving disc galaxies with preferentially prograde orbits \citep{Struck2012,Duc2013}.  The shape of tidal tails may be influenced by the shape of the gravitational potential wells and underlying dark matter haloes. \citet{Dubinski1996} showed through simulations that galaxies with a low halo-to-disc mass ratio can produce long tidal tails in the merging process.  \citet{Springel1999} further pointed out that long tidal tails can be generated when the halo spin in sufficiently large for merging galaxies with a high halo-to-disc ratio.  
Moreover, using TDGs on tidal tails,  \citet{Bournaud2003} found that dark matter haloes around spiral galaxies are at least ten times larger than the stellar discs. Therefore, long tidal tails are useful probes for dark matter halo parameters and galaxy structures.
The characteristics of tidal tails are affected by the orbital parameters, disc orientations as well as stellar and gas masses and the specific angular momentum of the merging galaxies \citep{Struck2012, Ploeckinger2018}.  Long tidal tails from mergers of disc galaxies thus provide hints on the orientation between their spin and orbital parameters. 

Moreover, star clusters and tidal dwarf galaxies (TDGs) can be formed in tidal tails \citep{BarnesHernquist1992, Bournaud2006,Wetzstein2007}.  Previous observational studies of TDGs focused on individual objects or small samples \citep{Duc2000,Duc1998,Hancock2009, Mohamed2011, Kaviraj2012}. A systematic investigation of tidal tails using a sufficiently large sample of merging galaxies is required to address how the tidal tails are created in reality,  draw constraints on how spiral galaxies merge with each other in the cosmic web,  and examine the formation of TDGs and star formation under these extreme conditions. 
 
Although there have been many simulations to study the morphology of merging galaxies and tidal tails.  However, it is not clear how many shapes real tidal tails have, how do they relate to the merger parameters and how they are formed. At the same time, solving the problem of the relationship between the TDG formed in the tidal tail and the shape of the tidal tail will help understanding merging processes and star formation in tidal tails.
 
In this work, we analyse a sample of 461 merging galaxies with long tidal tails (defined as having sizes comparable to the effective radii of the host galaxy, \citealt{Wen2016}) to characterize  tidal tails as well as their parent galaxies.  We briefly introduce the sample and data in Section~\ref{sec:data}. Section~\ref{sec:results} presents our analysis and results and we discuss the implications in Section~\ref{sec:discussion}.  Through out this paper, we adopt a concordance cosmology of $H_{0}=70$ km\,s$^{-1}$\,Mpc$^{-1}$ , $\Omega  _{\rm m}=0.3$ and $ \Omega_\Lambda =0.7$.  All photometric magnitudes are given in the AB system.

\section{ SAMPLE AND DATA} \label{sec:data}

We adopt the sample of 461 long-tidal-tail merging galaxies (LTTGs) in the COSMOS field from \citet{Wen2016}.  They identified these LTTGs using {\it Hubble Space Telescope} ({\it HST})/ACS F814W ($I$) imaging data and multi-band deep survey data and catalogs in COSMOS \citep[e.g.][]{Leauthaud2007,Capak2007}.  The {\it HST}/ACS F814W science images have a  scale of  0$\farcs$03 per pixel and reach a limiting magnitude limit of 25.6 mag for extended sources within a circular aperture radius of 0$\farcs$3.  The  Full Width at Half Maximum (FWHM) of the point-spread function (PSF) of the F814W images is 0$\farcs$09, corresponding to a physical size of 297\,pc at $z=0.2$ and of 723\,pc at $z=1$ \citep{Koekemoer2007}.  The {\it HST} imaging data cover an area of 1.64 deg$^2$ and provide high-resolution morphological information in the rest-frame optical for galaxies at $z<1$.  The stellar mass and photometric redshift catalogs come from \citet{Muzzin2013},  based on the UltraVISTA survey, providing a total of 154,803 $K_{\rm s}$-band-detected sources down to  5\,$\sigma=23.4$\,mag with a 90\,per\,cent completeness for point sources. 

In \citet{Wen2016}, a morphological classification method, namely the $A_{\rm O}$-$D_{\rm O}$ method, was used to select LTTGs.  As a non-parametric method for galaxy morphology, the $A_{\rm O}$-$D_{\rm O}$ method is dedicated to probing asymmetric  structures in the outskirts of a galaxy \citep{Wen2014}.   In the method, the image of a galaxy  is divided into two parts, the inner half-light region (IHR) and the outer half-light region (OHR).  Each region contains half the total flux of the galaxy. Two parameters are used: the outer asymmetry, $A_{\rm O}$, measures the asymmetry of OHR and  the outer centroid deviation, $D_{\rm O}$, measures the deviation (or offset) between the flux-weighted centroids of the IHR and the OHR.  The parameters are sensitive to the asymmetric structures in the outskirts such as long tidal tails. Galaxies with more disturbed morphologies  tend to have higher $A_{\rm O}$ and $D_{\rm O}$.  In practice, $A_{\rm O}$ is calculated using the formula:
\begin{equation}
         A_{\rm O}=\frac{\sum{|I_{0}-I_{180}|}-\delta_{2}}{ \sum{|I_{0}|}-\delta_{1}},
\end{equation}
where $I_0$ refers to the light distribution of OHR of an image; $I_{180}$ represents the $180^\circ$ -rotation version of $I_0$; $I_0$-$I_{180}$ yields the residual image; $\delta_{1}$ and $\delta_{2}$ are corrections for the noise contributions to the flux image $I_0$ and the residual image $I_0$-$I_{180}$, respectively.
The outer centroid deviation $D_{\rm O}$ is measured as below:
\begin{equation}
D_{\rm O}=\frac{\sqrt{(x_{\rm O}-x_{\rm I})^2+(y_{\rm O}-y_{\rm I})^2}}{R_{\mathrm  eff}},
\end{equation}
where ($x_{\rm I}$, $y_{\rm I}$) and ($x_{\rm O}$,$y_{\rm O}$) refer to the flux-weighted centroid position of  the IHR and of the OHR in pixels, respectively, and  $R_{\mathrm eff}$ is the normalized effective radius in pixels estimated using $R_{\mathrm eff}=\sqrt{n_c/\pi}$. Here $n_c$ is the area in pixels of the IHR. More details about the method can be found in \citet{Wen2014}. 

Two steps were adopted for identifying LTTGs in \citet{Wen2016}.  Firstly, they pre-selected 13,227 galaxies with morphologies showing a certain degree of disturbance using the  $A_{\rm O}$-$D_{\rm O}$  method from 35,076 galaxies with $\log(M_\ast/\rm M_\odot)\geq9.5$ within $0.2\le z\le1$ in the COSMOS field.  Secondly, the 13,227 candidates  were visually examined  using the {\it HST}/ACS F814W science images.  A total of 461 merging galaxies were identified with long tidal tails. See \citet{Wen2016} for more details about the sample selection.  We use this sample to analyse their morphologies and the substructures of their tidal tails.

\begin{figure}
         \centering
	{\includegraphics[width=0.95\columnwidth]{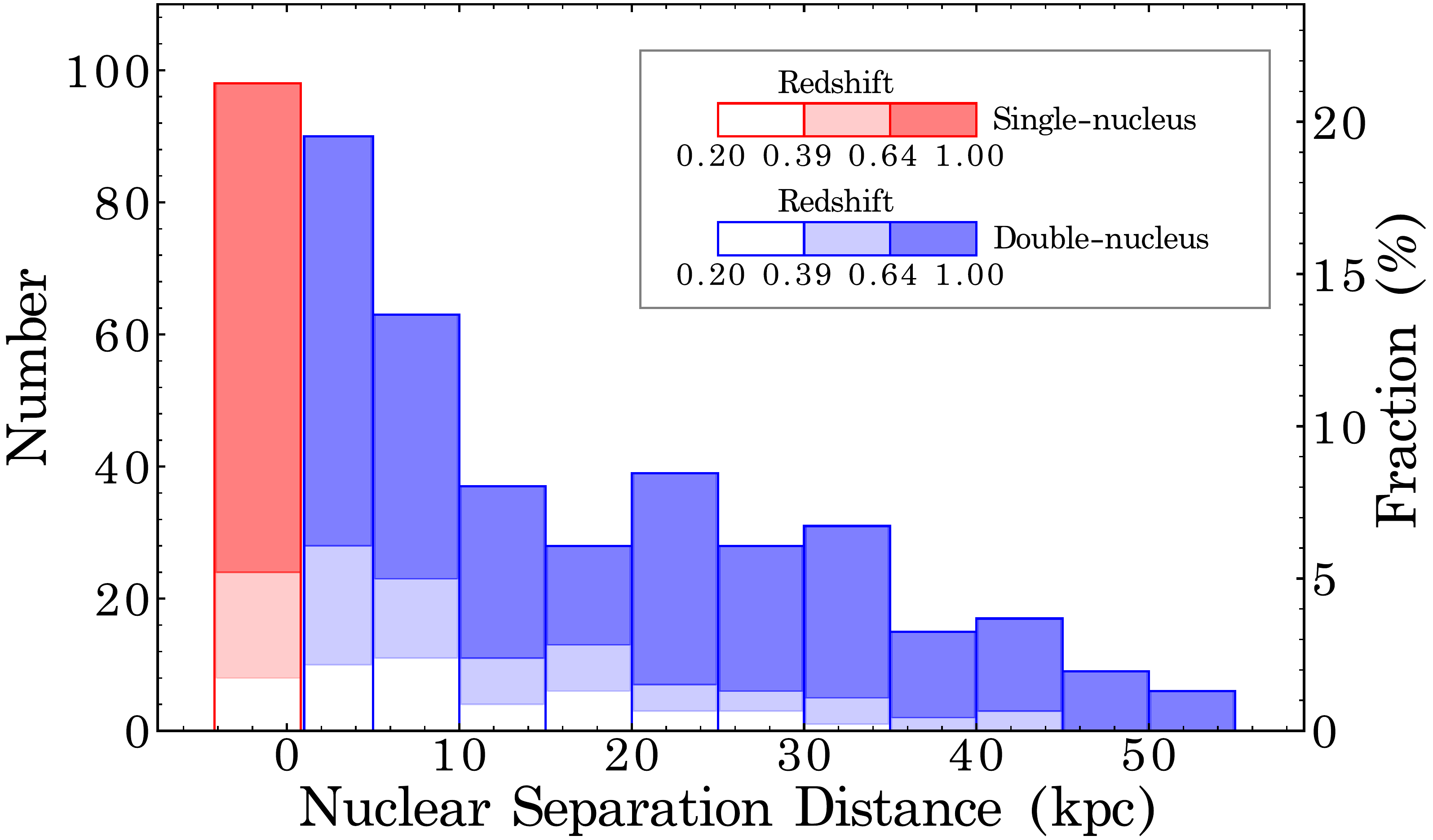}}
        \caption{Distribution of nuclear separation distance in our sample merging galaxies.  The distance set as zero is for galaxies with a single nucleus, as shown in the red bar. The blue bars show the distribution for two-nucleus galaxies. For the systems with more than two nuclei, the smallest separation distance is adopted. The blank, half-shaded and filled regions represent contributions from $0.2<z<0.39$, $0.39<z<0.64$ and $0.64<z<1.0$ respectively, so that they correspond to the same time interval of 1.8 Gyr.
 }        
      \label{fig:nsp}
\end{figure}

\section{Analysis and Results} \label{sec:results}

Using {\it HST}/ACS  F814W imaging data, we characterize the morphological properties of our sample LTTGs, including the nuclear separation distance, the shape and length of tidal tails and substructures of tidal tails such as clumps, TDGs and gaps.

\subsection{Nuclear Separation Distance of Merging Galaxies}

Simulations show that the growth of tidal tails is closely coupled with the orbital parameters and the  merging stage of a galaxy merger \citep{Toomre1972,Barnes1992}.  In a general merging process, two galaxies approach each other, pass their pericentre, and undergo a damped oscillation  with dynamical friction allowing the galaxies to lose energy and angular momentum,  while violent relaxation happening in the final stages as they merge into a single remnant. A close passage of one galaxy next to the other would lead to varying gravitational forces across the extended galaxies. The sides of the galaxies facing each other are more attracted than the opposite sides \citep{Duc2013}. This effect allows galaxies to produce a long tidal tail and a short counter tail during the merging process.  The nuclear separation distance is a crude tracer of the merging stage, although the separation oscillates a bit before the final collision.  In particular, after the first pass the nuclear separation distance increases but the two galaxies appear different from the approaching pair before the first pass. The merging galaxies after the first pass are strongly disturbed and thus can be distinguished from the galaxy pair in the beginning of merging without apparent tidal features (e.g., tidal tails). 

If tidal tails are associated with the remnant of a previous merger event,  the nuclear separation distance in the ongoing merger is irrelevant with the status of the tidal tails. We argue that such cases should be rare and have marginal effects on our analysis, considering that massive galaxies statistically underwent even less than one major merger event over 8\,Gyrs \citep[e.g.,][]{Jogee2009}.

\begin{figure}
\fbox{\includegraphics[width=0.31\columnwidth]{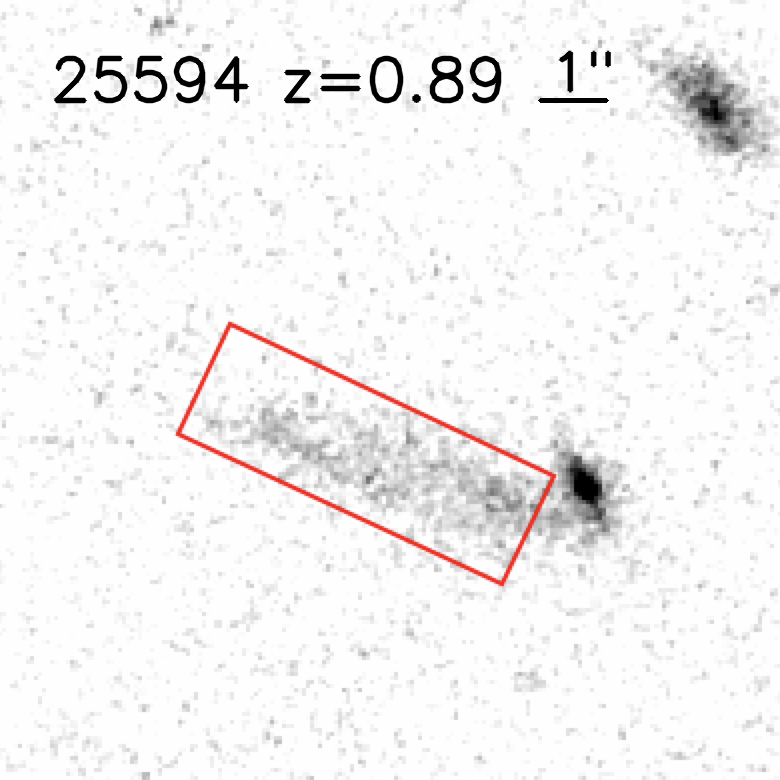}
	\includegraphics[width=0.31\columnwidth]{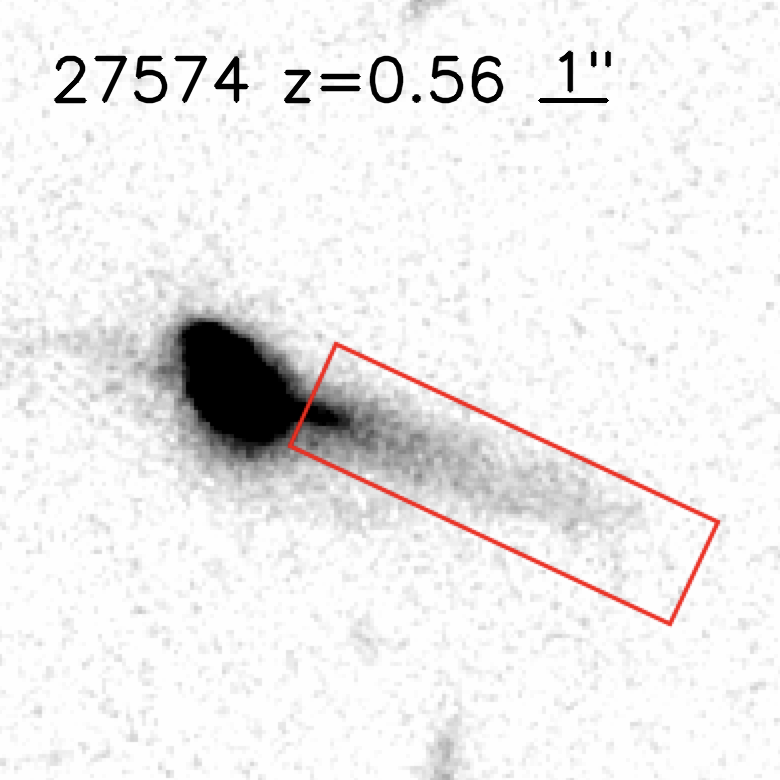}
	\includegraphics[width=0.31\columnwidth]{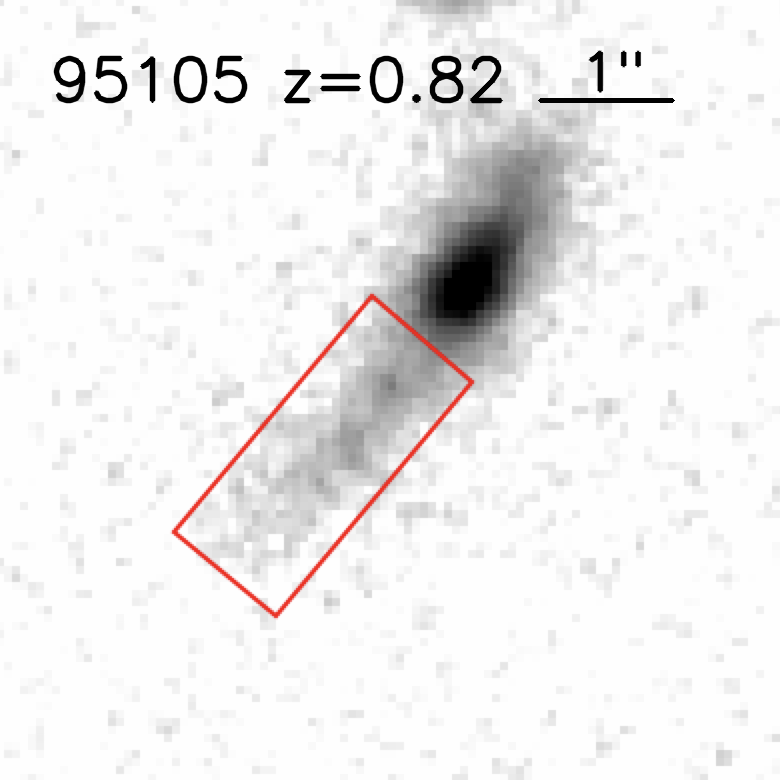} } \\
\fbox{\includegraphics[width=0.31\columnwidth]{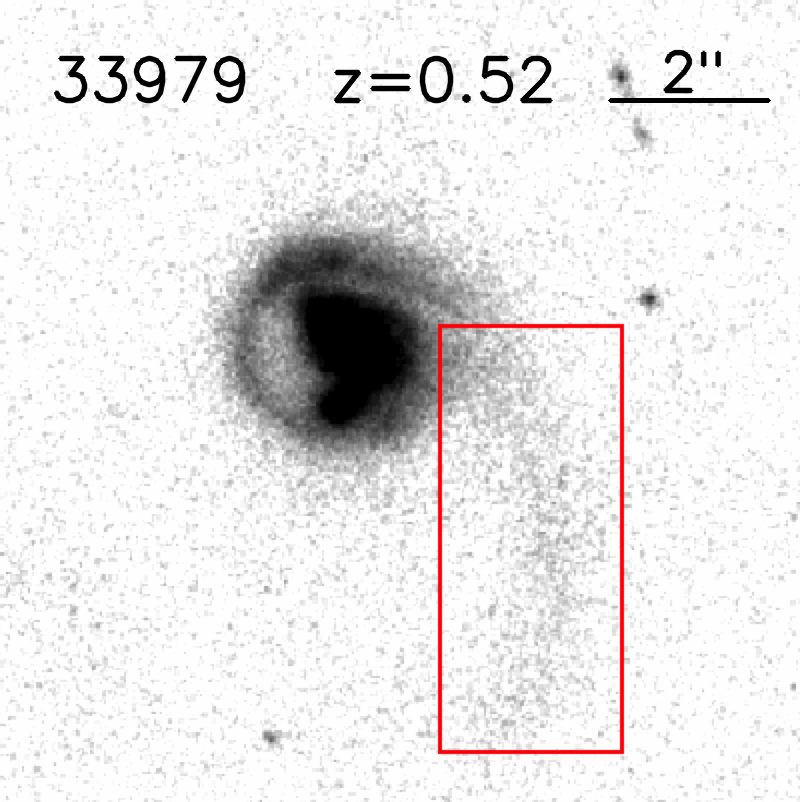}
	\includegraphics[width=0.31\columnwidth]{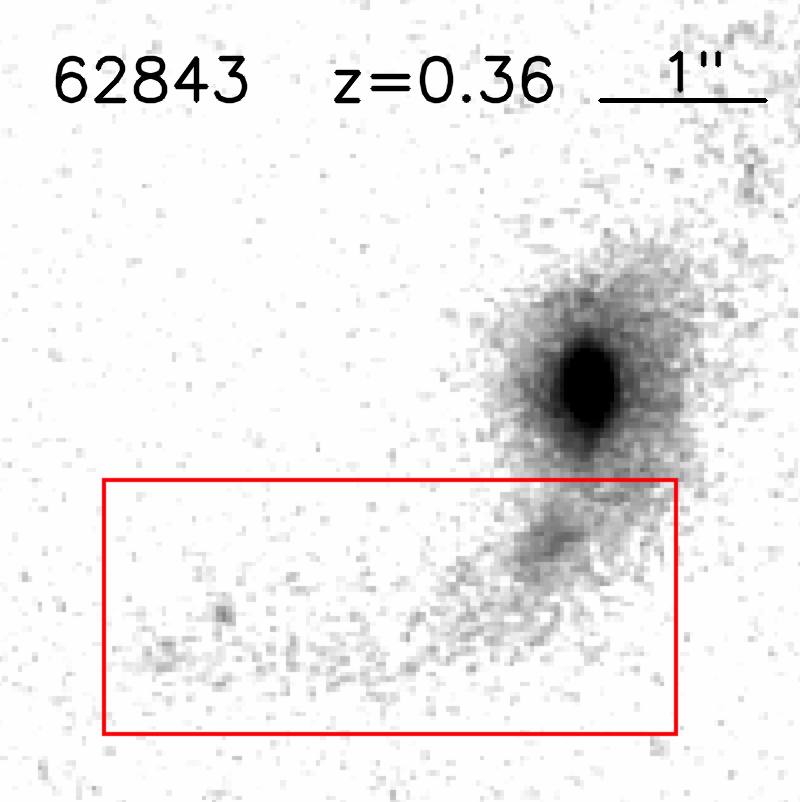}
	\includegraphics[width=0.31\columnwidth]{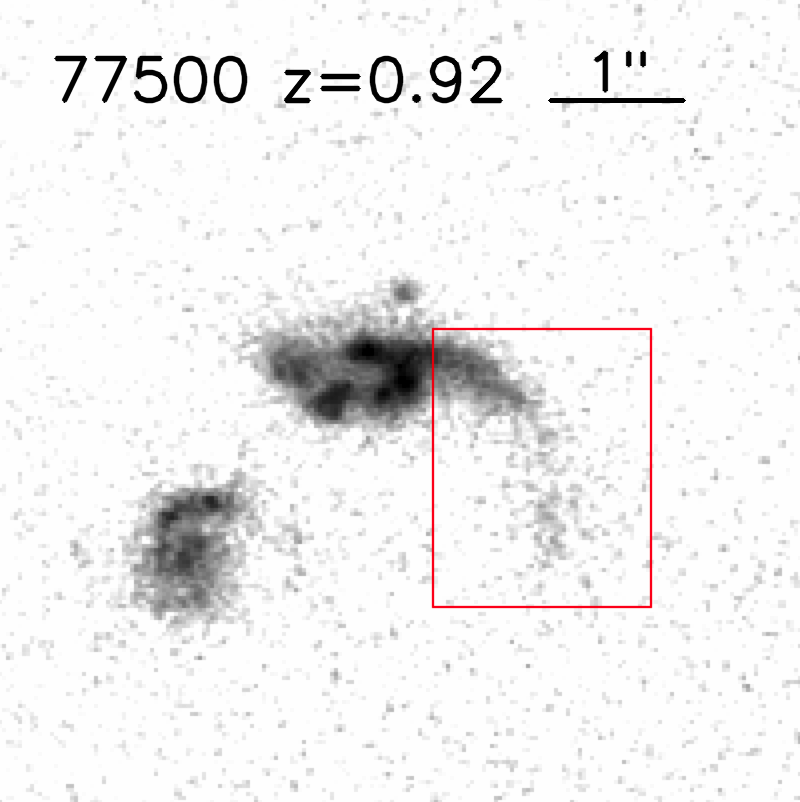} } \\
\fbox{\includegraphics[width=0.313\columnwidth]{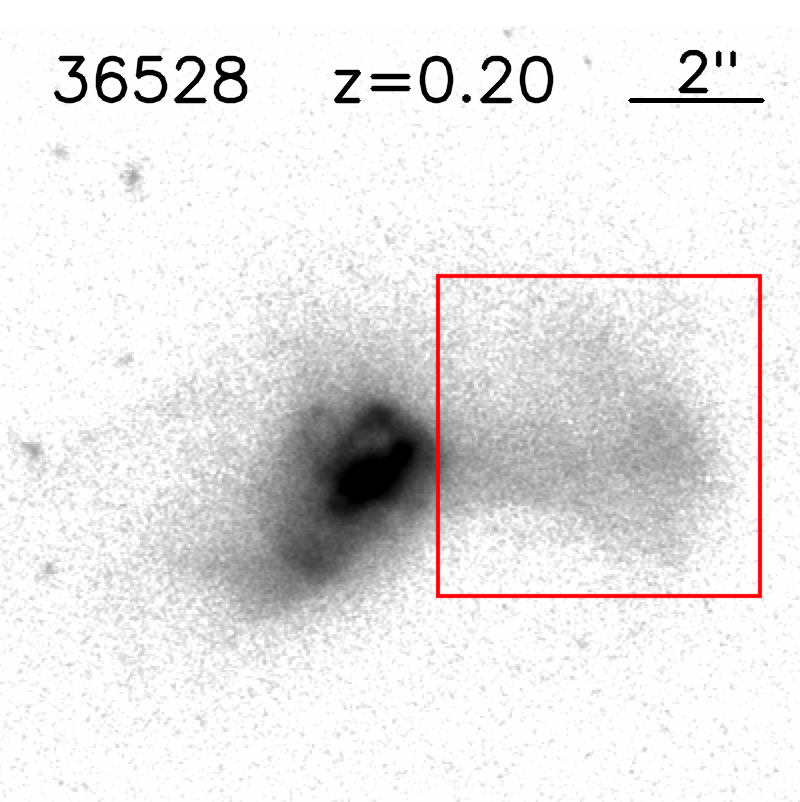} 
	\includegraphics[width=0.313\columnwidth]{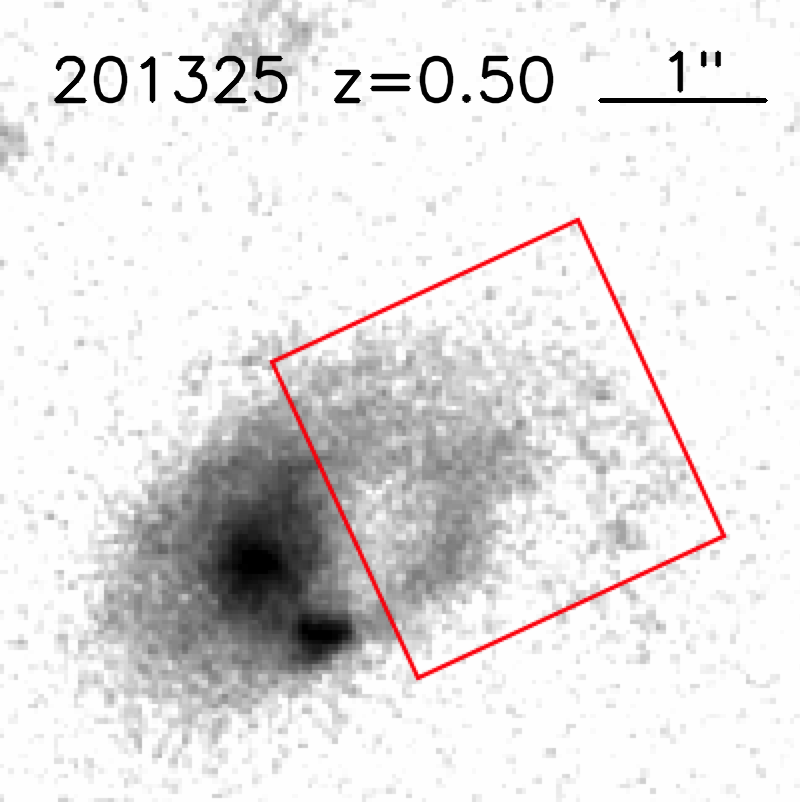}
	\includegraphics[width=0.313\columnwidth]{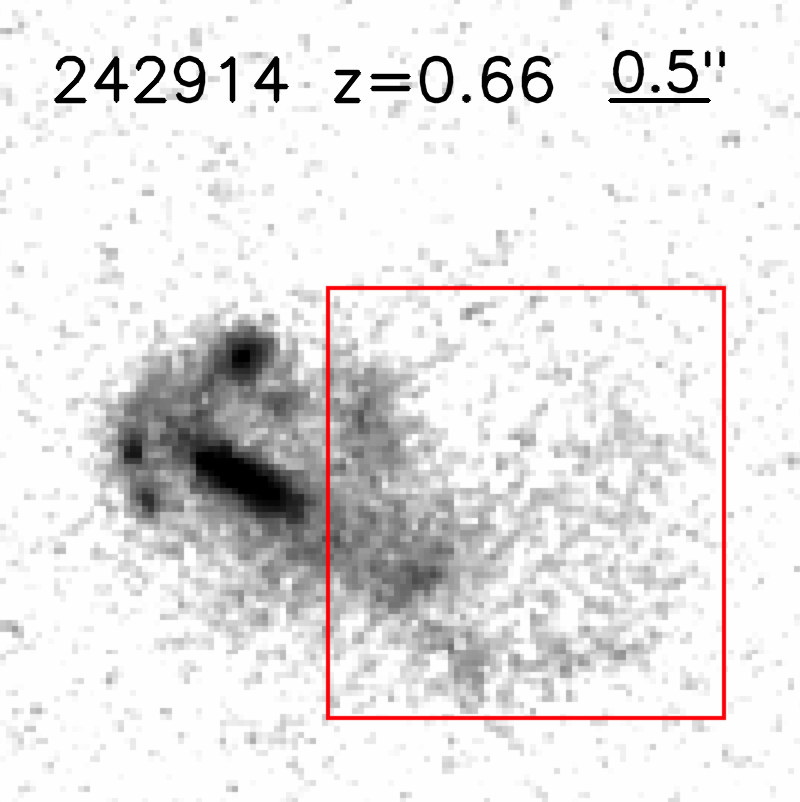}}
         \caption{Example {\it HST} F814W  images  of three types of tidal tails in our sample:   straight (top), curved (middle) and plume (bottom). The image size, target ID and redshift are labelled on each image. Red boxes mark the tidal tails.}
  \label{fig:tailtype}
\end{figure}

 We measure the nuclear separation distance ($d$) from the {\it HST} images of our sample galaxies.   
 For galaxy pairs, it is easy to find the centre of merging galaxies. For galaxies in their late merger phase, we use the method from \citet{Cui2001} to find galactic nuclei. We visually examine the light distribution of a merging system to find the core of the merging galaxies. If there is a dust lane, the single nucleus of a galaxy could be mistaken as a double-nucleus galaxy. In this sample, dust lanes are found in only a few merging galaxies in the final stage. If a galactic core is obscured by a dust lane, then we count it as a single core. Note that some sample galaxies contain only a single galactic nucleus, being most likely in the final stage of the merging process. We set $d=0$ for these single-nucleus mergers. In contrast,  two nuclei are distinguishable even in a merger remnant with disturbed morphology. We refer such systems to as merger remnants with $d<5$\,kpc. On the other hand,   8\,per\,cent of our sample merging systems contain more than two galaxies (or galactic nuclei) with $d<30$\,kpc. For  these multiple-galaxy merging systems, the smallest separation distance to the galaxy with long tidal tails is taken as the nuclear separation distance.

Figure~\ref{fig:nsp} shows the distribution of the measured nuclear separation distance for our 461 sample LTTGs.  The uncertainties of our measurements of the nuclear separation distance are typically 10\,per\,cent, and increase at decreasing separation. The typical phot-$z$ error ($\delta z/(1+z)\sim0.013$ at $z<1.5$)  induces negligible  uncertainties in size measurement (0.05\,kpc in the transverse distance at 1" at $z=1$).  Our sample consists of  98  (21\,per\,cent) single-nucleus mergers, and 90 (20\,per\,cent) mergers with $d<5$\,kpc, 128 (28\, per\,cent) close pairs with $5\leq d\leq 20$\,kpc,  and 145 (31\,per\,cent) galaxy pairs with $d>20$\,kpc.   It is clear that the long tidal tails are mostly associated with the merging galaxies in the later stages ($d<20$\,kpc). This is not surprising because long tidal tails are caused by strong interactions. For the galaxy pairs with $20<d<55$\,kpc, we suspect that these galaxy pairs have undergone the first pass as the long tidal tails can be seen as a sign of violent interaction. 
Simulations indicate that long tidal tails can be visible over $\sim$1\,Gyr, mostly between the first pass and the final coalescence  in the observations similar to these adopted here \citep{Barnes1992,Lotz2008b,Ji2014}. Roughly speaking, the phase with $d>20$\,kpc accounts for 50\,per\,cent of the time-scale, while the phase with $d<5$\,kpc represents 35\,per\,cent \citep{Ji2014}.  Considering the projection effect, a large fraction of widely-separated merger pairs ($d>20$\,kpc) may appear as  $5\leq d\leq 20$\,kpc. The three bins of $d<5$\,kpc, $5\leq d\leq 20$\,kpc and $d>20$\,kpc account for roughly 30, 30 and 40\,per\,cent of the long-tail-visible time-scale. We thus conclude that the results shown in Figure~\ref{fig:nsp}  are globally consistent with the time-scales given by the simulations, although the uncertainties of time-scales are large.

\begin{figure}
	\centering
	\includegraphics[width=0.95\columnwidth]{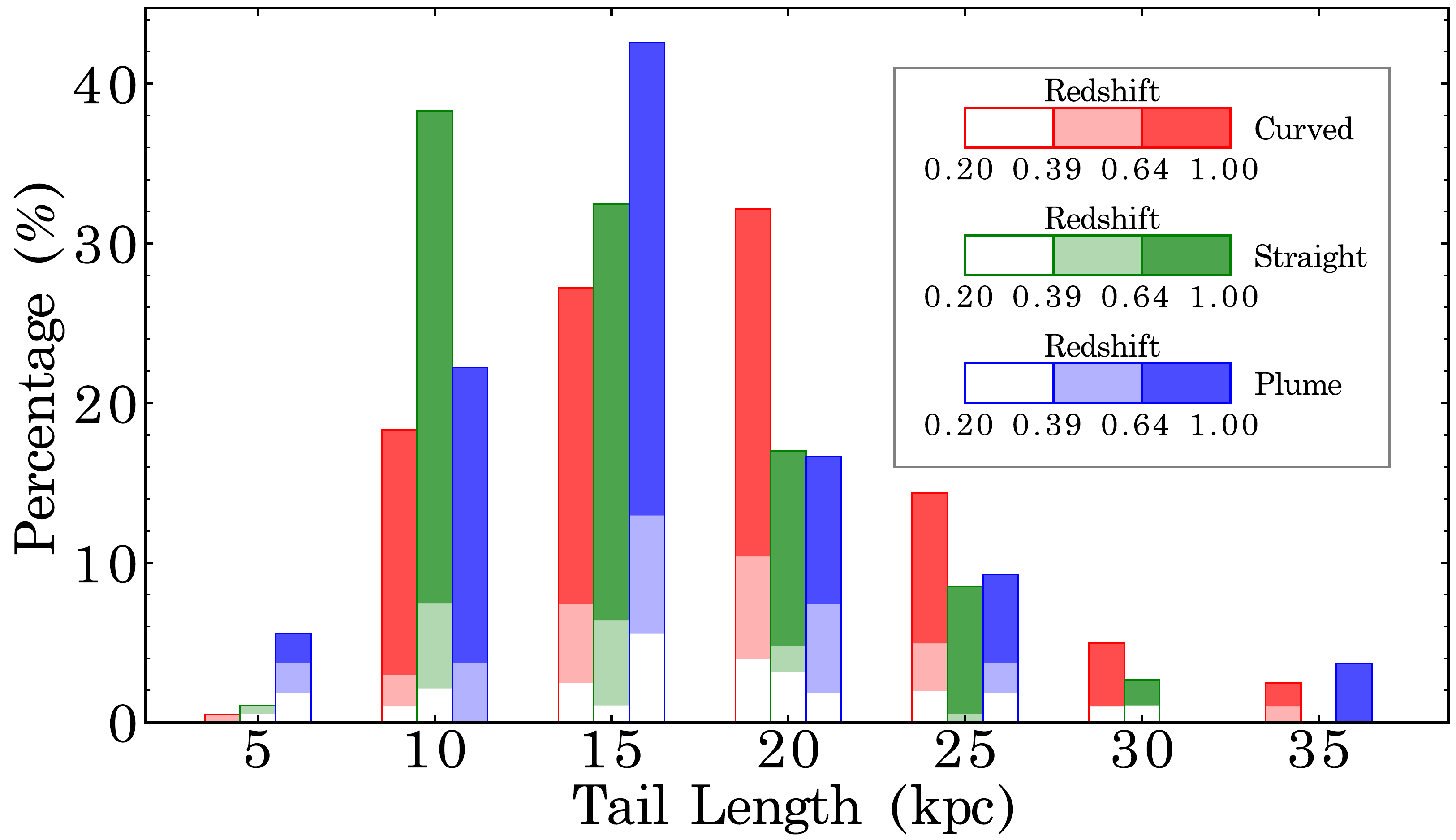}
         \caption{Histogram of tail lengths for three types of tidal tails. The total number is 192, 213 and 56 for straight, curved and plume, respectively.  The data bin width is 5\,kpc and data bars are placed at the  centre of corresponding data bins. Red and blue bars are shifted for clarification. The blank, half-shaded and filled regions represent contributions from $0.2<z<0.39$, $0.39<z<0.64$ and $0.64<z<1.0$ respectively, so that they correspond to the same time interval of 1.8 Gyr.}
         \label{fig:length}
\end{figure}

\subsection{ The Shape of Tidal Tails}

We visually examine the shapes of long tidal tails in our sample galaxies. A pair of tidal tails is often generated by one disc galaxy while merging. Due to the projection effect and low surface brightness of counter tails, only one tidal tail is visible in some cases.  Moreover, we note that the shape of a tidal tail seen from the {\it HST} image is not the same as the configuration in the real space.  Three of us (J.R., M.Q. and R.W.) independently classified the morphologies of the long tidal tails. The agreed classification by at least two of us is adopted. Controversial results were obtained for about 10\, per\,cent targets. We were re-examined to set classification. Our visual examination of  all long tidal tails in our sample shows that these long tidal tails can be roughly classified into three shapes.

Figure~\ref{fig:tailtype} presents example images of the three types of tidal tails, defined as straight, curved, and plume.  The key features of the three shape types are as follows:
\begin{description}
	\item[\bf Straight] --- the tidal tails are straight or slightly curved. The width of the tails is usually significantly smaller than their length and the size of their host galaxies.
	\item[\bf Curved] ---  the tidal tails are  strongly curved with a shape like ``S'', ``U'' or ``V''.  The tails are relatively thin and their width is much smaller than their length.  We also include ring-shape tails into this type. 
	\item[\bf Plume] --- the tidal tails spread widely in a form without a certain figure.  Such tails usually  have low  surface brightness (SB), sticking to the outskirts of the host galaxies. 
\end{description}
In  interacting galaxies with two tidal tails,  one is a long tail and the other is usually shorter. The long tail is often curved or straight, and the counter tail sometimes appears to be a plume tail.  In our classification, we count a merging system with a plume tail as the plume type if the plume tail is the major one or comparable to the other tail in size. If there is no plume tail, we use the longest tail to identify the shape type, even though the shorter tails might have a different type.

\begin{figure}
	\fbox{\includegraphics[width=0.46\columnwidth]{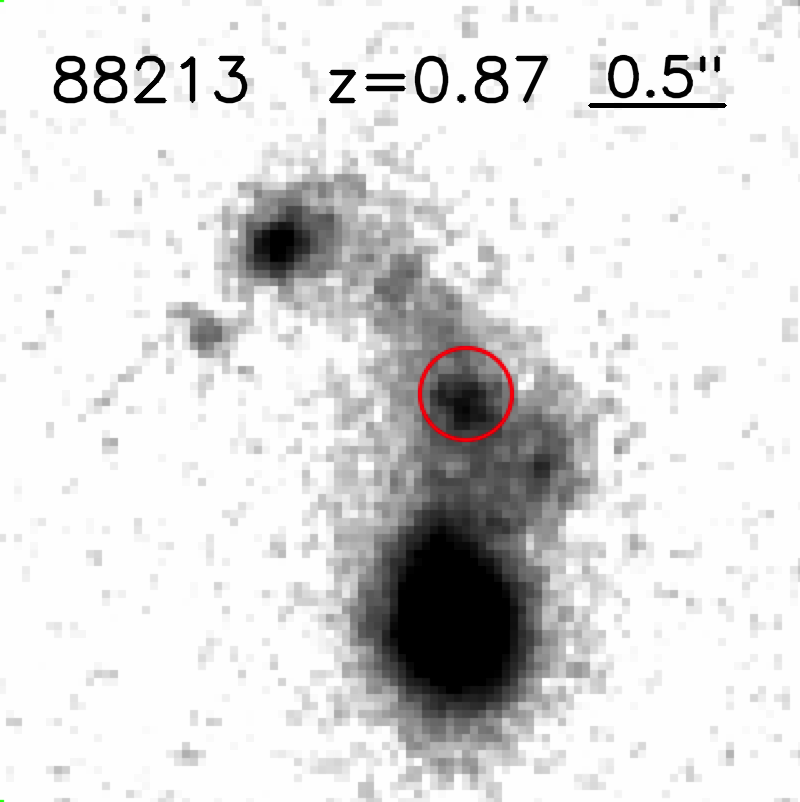}}
	\fbox{\includegraphics[width=0.46\columnwidth]{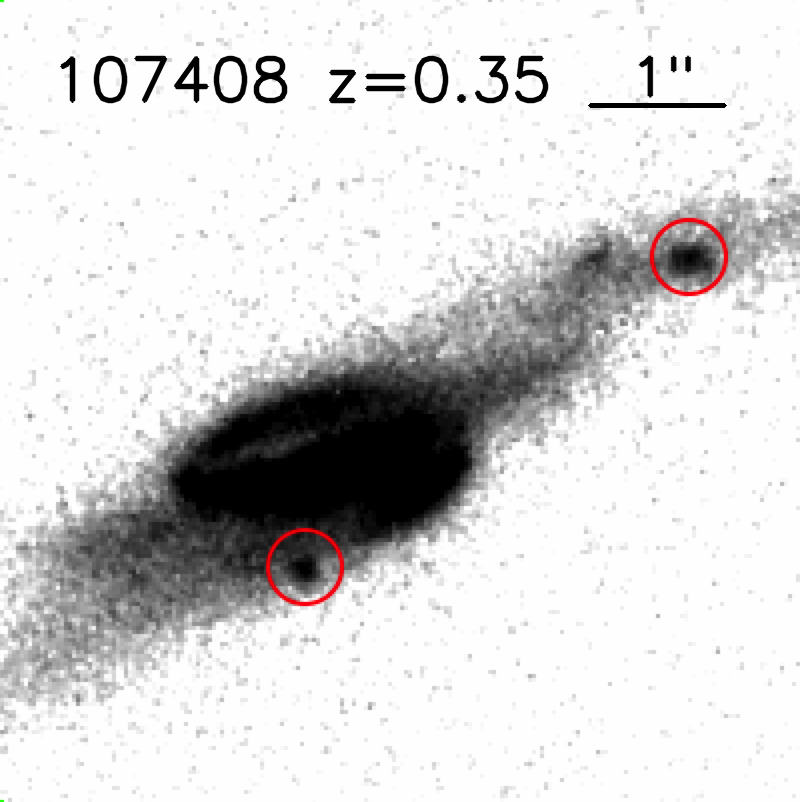}}	\\ 
	\fbox{\includegraphics[width=0.46\columnwidth]{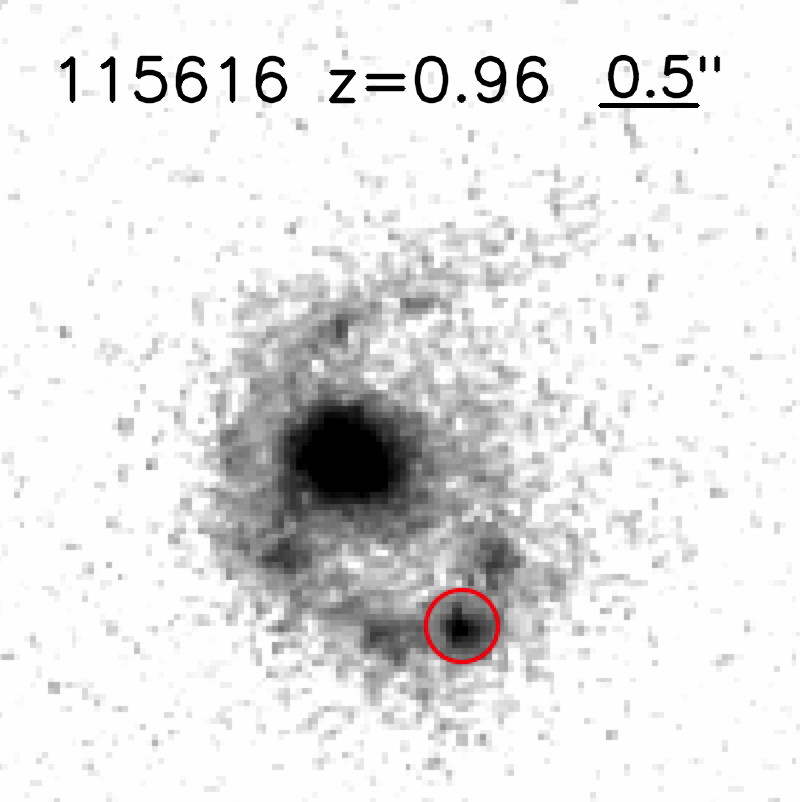}}
	\fbox{\includegraphics[width=0.46\columnwidth]{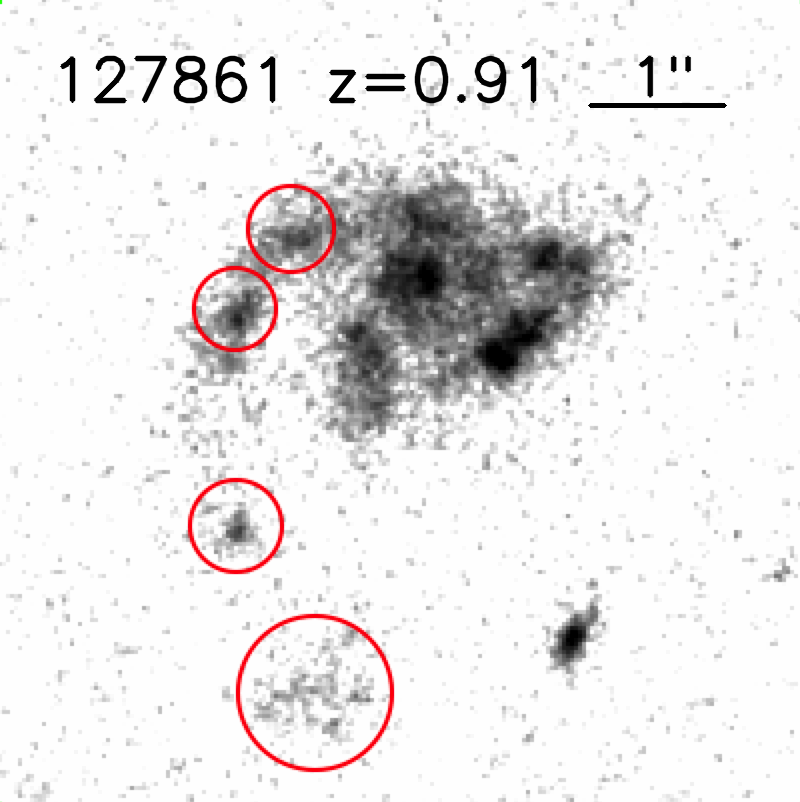}}
         \caption{Clumps on the tidal tails in four LTTGs from our sample. The red circles mark the identified clumps.} 
         \label{fig:clump}
\end{figure}

The number of our sample LTTGs with straight, curved or plume tails is 192 (41\,per\,cent), 213 (47\,per\,cent) and 56 (12\,per\,cent), respectively. It is clear that the long tidal tails in our sample mostly have straight and curved shapes. The plume tails are relatively rare.  We notice that the number of the straight tails is comparable to that of the curved tails. This is inconsistent with the idea that {\it the straight tails are the projected curved tails} as curved tails seen edge-on should account for only a small fraction of all viewing angles.  A quantitative analysis on the separation of the two kinds of straight tails would need a larger sample limited in stellar mass and redshift in order to eliminate various effects.  The future wide and deep imaging surveys (e.g., EUCLID, \citealt{Laureijs2011}; China Space Station Telescope, \citealt{Gong2019}; and MESSIER, \citealt{Valls-Gabaud2017}) will provide sufficiently large data sets to address this issue. 

We measure the length of tidal tails based on the optical brightness of  the stellar component of a tail. The H\,I gas of the tail may extend to much longer distances  \citep[e.g.,][]{Duc2000}. A  surface brightness threshold to 0.5\,$\sigma$ of background noise level is adopted to draw the boundaries of the tail.  For the straight tails, we enclose the entire tail within a rectangle, and the length of the long side of the rectangle is taken as the length of the tail. For the curved tails, we draw a curved line along the ridge of the tail and measure its length.  For plume tails, we use an ellipse to fit the boundaries of a tail and adopt as the length of the tail the major axis of the ellipse.
 We point out that the cosmic dimming effect may influence the tail length measurements.  Our sample LTTGs span over a wide redshift range of $0.2<z<1$ and tail lengths tend to be underestimated at higher $z$.  We will address this issue in Section~\ref{sec:333}.

\begin{figure}
          \fbox{\includegraphics[width=0.46\columnwidth]{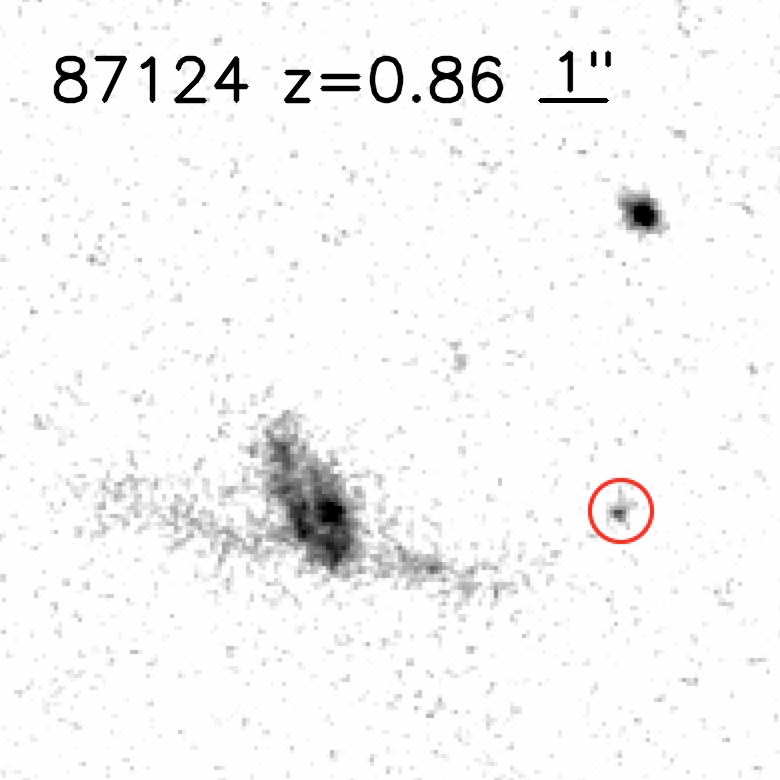}}
          \fbox{\includegraphics[width=0.46\columnwidth]{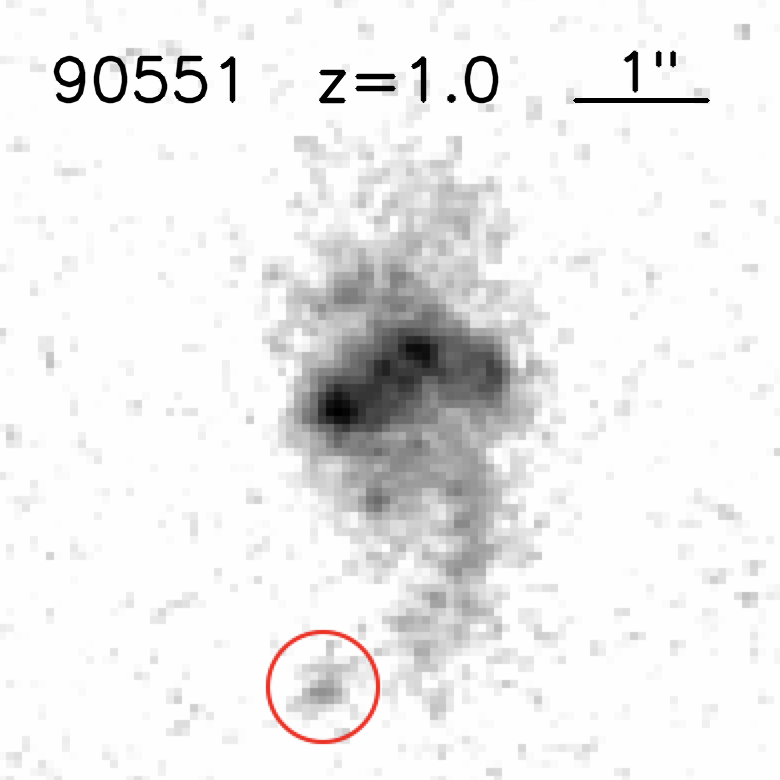}} \\
          \fbox{\includegraphics[width=0.46\columnwidth]{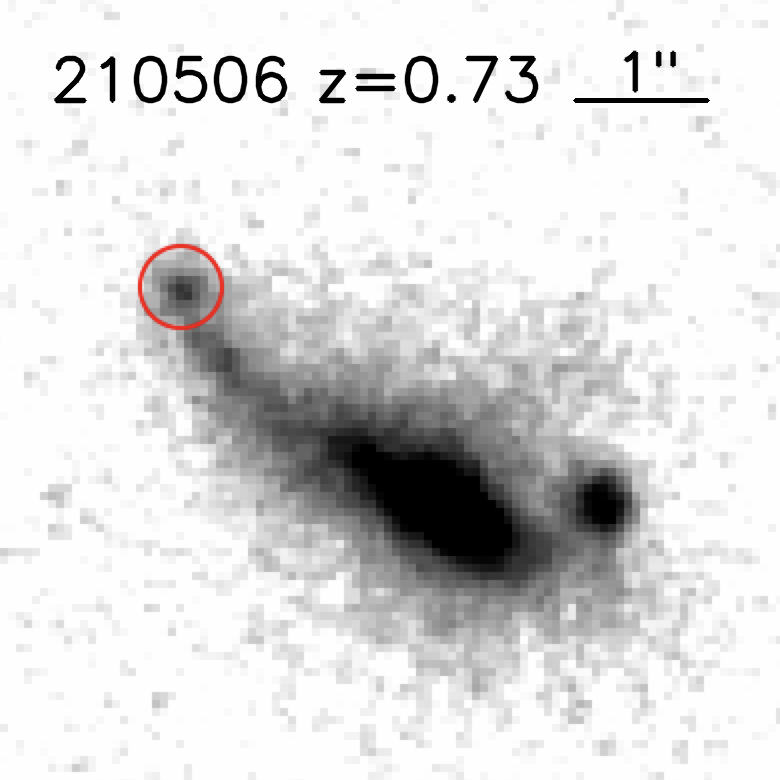}} 
          \fbox{\includegraphics[width=0.46\columnwidth]{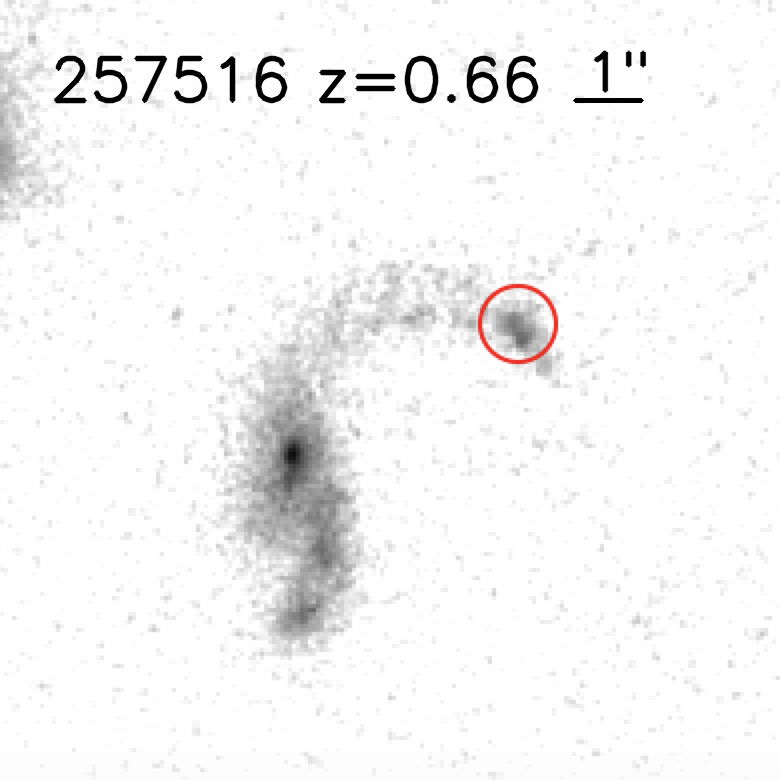}}
         \caption{{\it HST}/ACS F814W images of four tidal dwarf galaxies identified at the tip of long tidal tails. The red circles mark the massive TDGs at the tip of the tidal tails.}
         \label{fig:tdgs}
\end{figure}

 Figure~\ref{fig:length} shows the histogram of the tail length for the straight, curved and plume tails. We can see that the optically-detected tails in our sample mostly distribute in the length range of $7.5 - 22.5$\,kpc.  Assuming a rotating velocity of 200\,km\,s$^{-1}$, it would take $0.04 - 0.1$\,Gyr to produce these lengths, a time-scale much shorter than the typical merger time-scale of $\sim$1\,Gyr \citep{Lotz2008b,Ji2014}.   The three types of tails are similar in the length distribution, although the straight tails tend to be slightly shorter.


 \begin{table*}
 \centering
 \renewcommand\tabcolsep{5.0pt} 
  \caption{The properties and measured tail parameters of our sample of 461 LTTGs.}
  \label{tab:lttgs}
  \centering
  \begin{tabular}{lcccccccccc}
    \hline\hline
    ID & R.~A. & Dec & $z$ & $\log (\frac{M_{\ast}}{M_\odot})$ & Nucleus Number$^{\rm a}$ & Nuclear Distance & Shape Type$^{\rm b}$ & Tail Length & $N_{\rm TDG}^{\rm c}$  & LP$_{\rm TDG}^{\rm d}$ \\
        & (J2000.0) & (J2000.0) &   &  &  &  (kpc) &  & (kpc) & &  \\

    \hline
218	&	150.4096792	&	1.6409722	&	0.94	&	10.20	&	S	&	-	&	L	&	15.54	&	1 &	1	\\
708	&	150.3942292	&	1.6660750	&	0.60	&	9.66	&	D	&	2.52	&	L	&	4.46	&	0 & 	0	\\
737	&	150.6348000	&	1.6665639	&	0.85	&	10.10	&	S	&	-	&	C	&	17.50	&	1 & 	1	\\
939	&	150.5893583	&	1.6727472	&	1.00	&	10.40	&	M	&	3.40	&	C	&	11.04	&	0 & 	0	\\
1646	&	150.3790417	&	1.6888722	&	0.32	&	10.80	&	D	&	6.68	&	P	&	22.29	&	0 & 	0	\\
1658	&	150.6933292	&	1.6939667	&	0.97	&	10.70	&	D	&	13.23	&	L	&	21.22	&	0 &	 0	\\
1692	&	150.3774917	&	1.6955000	&	0.88	&	9.65	&	D	&	35.08	&	C	&	12.32	&	1 & 	1	\\
3042	&	150.4835500	&	1.7334778	&	0.71	&	10.30	&	M	&	13.51	&	C	&	32.39	&	0 & 	0	\\
3362	&	150.6746708	&	1.7463167	&	0.96	&	9.82	&	S	&	-	&	L	&	14.19	&	0 & 	0	\\
4335	&	150.4133292	&	1.7678472	&	0.91	&	9.79	&	D	&	11.07	&	P	&	12.15	&	0 & 	0	\\
... & ... & ... & ... & ... & ... & ... & ... & ... & ... \\
 \multicolumn{10}{l}{(The machine-readable and complete version of this table is available online.)} \\
 \hline\hline
 \multicolumn{10}{l}{
 $^{\rm a}$Nucleus Number --- S: single;  D: double;  M: multiple. 
 $^{\rm b}$Shape Type ---  L: straight,  C: curved, P: plume. 
 $^{\rm c}N_{\rm TDG}$ --  Number of TDGs on tidal tail.} \\
 \multicolumn{10}{l} {$^{\rm d}$LP$_{\rm TDG}$ -- Location Parameter, 0: no TDG; 1: TDGs at the tip of tidal tails; 2: TDGs on tidal tails; 3: TDGs at the tip and on the tidal tails. }
 \end{tabular} 
 \end{table*}

\subsection{Substructures of Tidal Tails}

Gravitational collapse drives gas to form clumps in a tidal tail. The clumps can form Super Star Clusters (SSCs) or TDGs within a relatively short time-scale.  In contrast, the stellar component in low  surface brightness along the tidal tail is not often bounded into these clumps. A detailed  investigation of the substructures of tidal tails enables to shed light into the formation of the clumps \citep{BarnesHernquist1992, Bournaud2006, Wetzstein2007}.

\subsubsection{Clumps}

We visually examine the substructures of all tidal tails and identify clumps formed on the tails. We identify a clump from the knot with the largest closed contour with encircled area large than 28 pixels$^2$ ($r$=FWHM) and the central surface brightness higher than 1.2 times the  surface brightness of the surrounding tidal tail \citep{Wen2012}. This criterion allows us to pick up clumps of PSF-corrected size $R_{\rm e}>0.2$\,kpc at $z=1$ and smaller clumps at $z=0.2$, indicating to our identified clumps are only complete at $R_{\rm e}>0.2$\,kpc over $0.2<z<1$. 
Figure~\ref{fig:clump} shows the examples of clumps identified on tidal tails of four merging galaxies. 
In total, we identify 177 clumps from our  sample of 461 LTTGs.  Clumps are often present in curved and straight  tails especially at the tip of tidal tails.  Plume tails usually have smooth surface brightness and barely contain clumps in this sample.

\begin{figure}
    \centering
    \includegraphics[width=0.95\columnwidth]{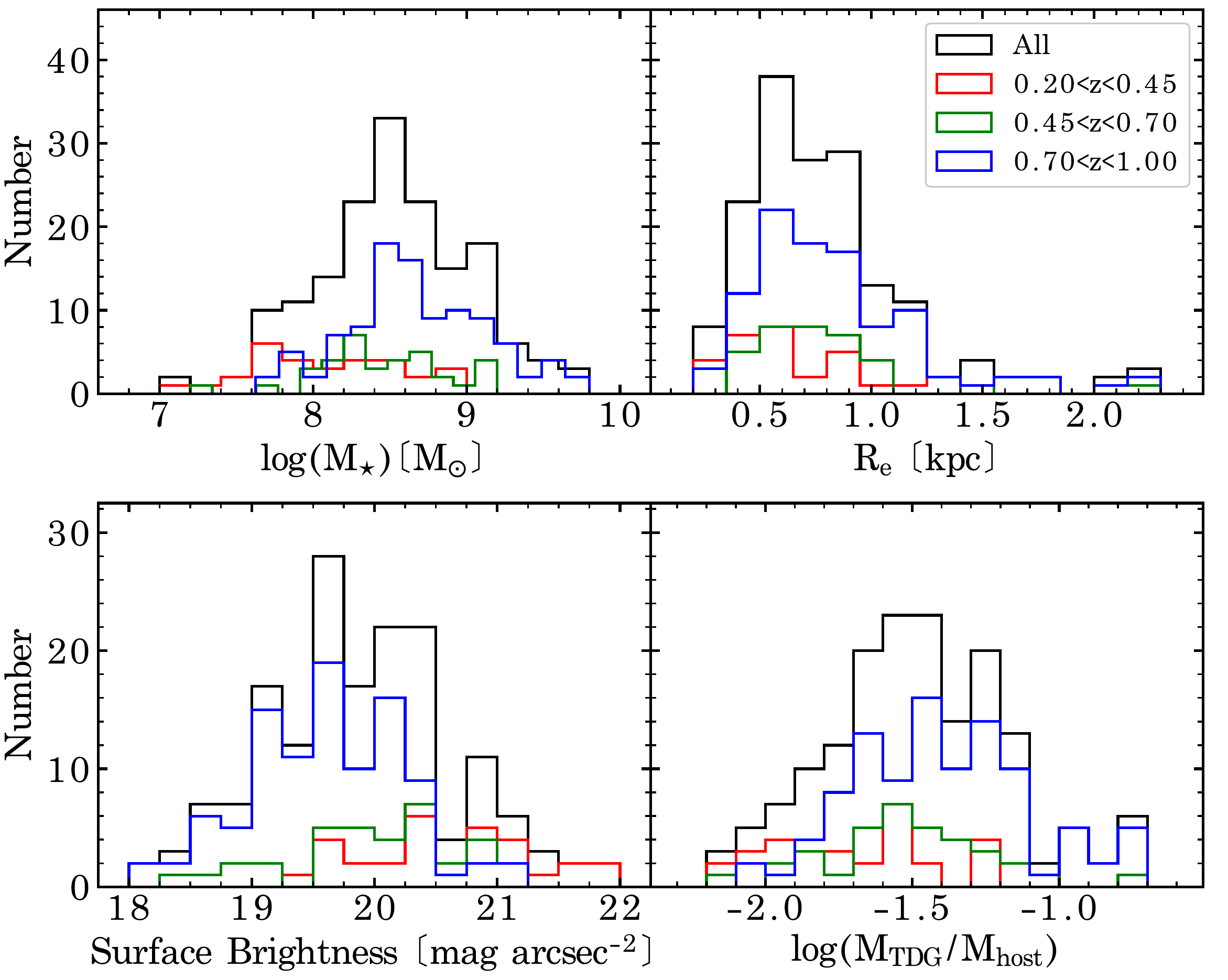}
    \caption{Distributions of all 165 TDGs in stellar mass, effective radius, the average rest-frame surface brightness within effective radius and TDG-to-host mass ratio.}
    \label{fig:tdg_dis}
\end{figure}

\begin{figure}
	\centering
	\includegraphics[width=0.95\columnwidth]{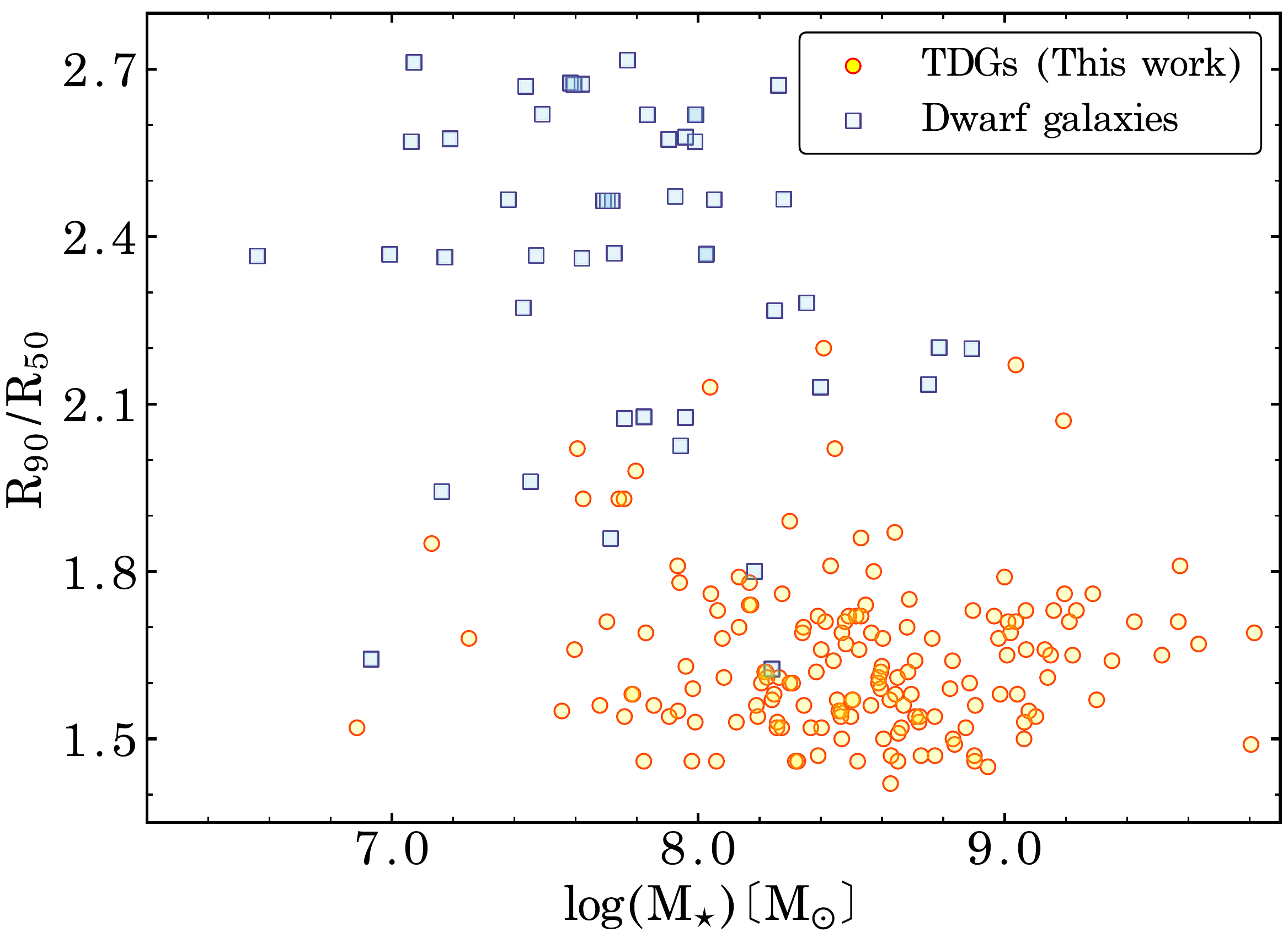}
        \caption{Stellar mass versus concentration parameter ($c=R_{90}/R_{50}$) for 165 TDGs identified from our LTTG sample (red circles) and 49 dwarf galaxies from \citet{Conselice2003a} (blue squares).}
      \label{fig:tdg_mc}
\end{figure}

It is challenging to determine the background in conducting photometry on a clump  and its boundaries from the host tidal tail.  The mean  surface brightness  of the surrounding area of the host tidal tail is estimated first and taken as the background for the clump photometry. At the tip of the tails, the background is directly derived from an annulus close to the clump. We use a set of evenly increasing apertures to measure fluxes for the clump and draw the growth curve to pinpoint the turnover radius where the curve turns flat.  The radial light profile is converted from the growth curve. The outer part of the profile, where the surface brightness is lower than that of the tidal tails, is fitted  with a power-law function $\Sigma _{\rm flux}$ \,=\,$\alpha r^{\beta}$\,+\,$\gamma$ and we derive the missing flux out of the turnover radius using the light-weighted best-fitting radial profile. The total flux and the half-light radius (i.e., effective radius $R_{\rm e}$) are derived from the composite radial profile (inner observed + outer best-fitting) of the clump  after correcting for the PSF effect using an empirical PSF radial profile derived from stars.  We also measure the radius within which 90\,per\,cent of light is enclosed from the radial profile.  The central  surface brightness, i.e., the mean  surface brightness within $R_{\rm e}$, is also calculated. We use the stellar mass-to-light ratio ($M/L$) of the host galaxy to estimate stellar mass for the clump. We caution that  this may lead to a slight overestimate of stellar mass as clumps usually have younger stellar populations than their host galaxies.  

Table~\ref{tab:lttgs} presents the catalog of 461 LTTGs and their measured tail parameters. The table columns include ID,  R.A., Dec, redshift, stellar mass, number of nuclei, nuclear separation distance, shape type of tidal tail, tail length, number of TDG, and the location parameter.

\subsubsection{TDGs}

 Within the sample of 177 clumps we identify TDGs. The confirmed TDGs in the literature \citep{Duc1998, Monreal-Ibero2007, Wen2012} show different properties from the dwarf galaxies in the Local Group \citep{Mateo1998}.  Following \citet{Wen2012}, we adopt three criteria to determine if a clump is a TDG:  1) stellar mass $M_\ast>10^7$\,M$_\odot$ ; 2) effective radius $R_{\rm e}>0.20$\,kpc ; 3) central  surface brightness brighter than 1.2 times that of the host tidal tails. Note that we relax the lower limit of $R_{\rm e}$ from 0.5\,kpc given in \citet{Wen2012} to 0.20\,kpc because we find some TDGs at the tip of tidal tails to be more concentrated. Applying  these criteria to all 177 clumps,  we obtain 165 as TDGs residing in 126 LTTGs over the redshift range $0.2< z<1$.  Four examples of TDGs identified at the tip of long tidal tails are shown in Figure~\ref{fig:tdgs}.  The identification of 165 TDGs from 461 LTTGs gives a TDG production rate 0.36 per merger.  We notice that TDGs can be found in tidal tails as well as at the tip of the tails. We introduce a parameter to show the location of TDGs in their host tails. The location parameter can be found in Table~\ref{tab:lttgs}.

\begin{figure}
         \centering
	\includegraphics[width=0.95\columnwidth]{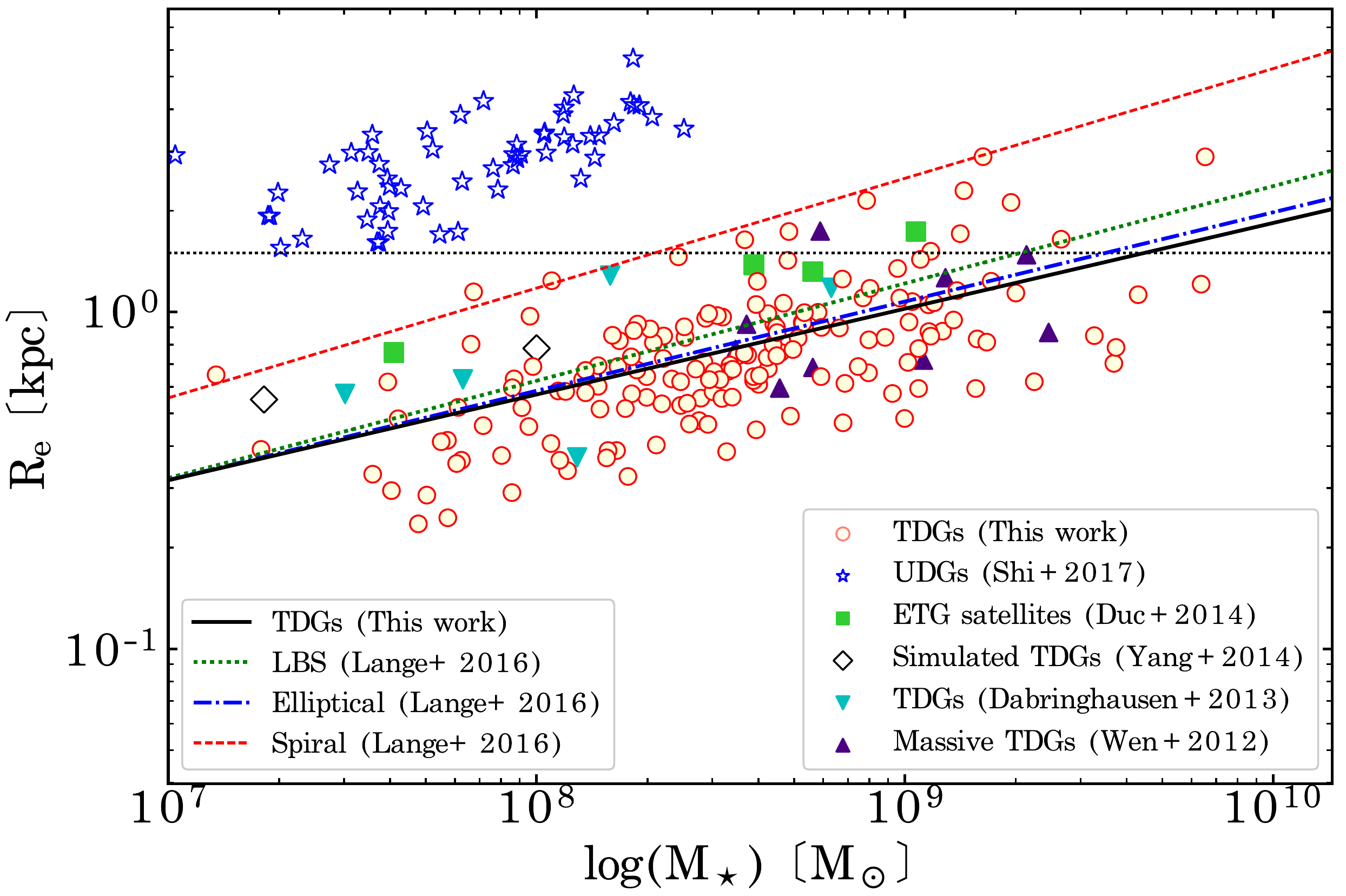}
	 \caption{ Stellar mass versus effective radius for TDGs and dwarf galaxy populations collected from the literature.  Red filled circles represent TDGs from our sample. Blue stars refer to ultra-diffuse galaxies from \citet{Shi2017}. Green filled squares are the satellites of early-type galaxies from \citet{Duc2014}. Cyan filled triangles show TDGs in the Local Group from \citet{Dabringhausen2013} and dark-purple triangles are TDGs at intermediate redshifts \citet{Wen2012}.  TDGs generated in simulations from \citet{Yang2014} are displayed with black open diamonds. The solid black line represent the best fit $M_*$ - $R_{\rm e}$  relation of TDGs in this work. The blue dot-dashed line, green dotted line and magenta dashed line represent respectively the $M_*$ - $R_{\rm e}$  relation of elliptical galaxies with $M_\ast<10^{10}$\,M$_\odot$, little blue spheroid (LSB) and spiral galaxies from \citet{Lange2016}.}
\label{fig:tdg_mr}
\end{figure}

Figure~\ref{fig:tdg_dis} presents the distribution of stellar mass, F814W central  SB, effective radius and TDG-to-host mass ratio for the 165 TDGs identified from our sample. The median of stellar mass is $ 3.4\times 10^8$\,M$_\odot$ and the average of $R_{\rm e}$ = 0.81\,kpc. Compared with TDGs in  \citet{Wen2012} and \citet{Duc2014}, our 165 TDGs are statistically more massive. This is not surprising because these TDGs are distributed in the redshift range of $0.2<z<1$.

TDGs are believed to form through gas collapse followed by infall of stars \citep{Duc1998,Duc2000} and new stellar populations can be formed at the centre of TDGs. The competition between collapse time-scale and star formation time-scale controls the light concentration of the TDGs.  We  quantify the concentration of TDGs using a concentration parameter $c$, defined as the ratio between the 90\,per\,cent-light and 50\,per\,cent-light radii ($c = R_{90}$/$R_{{50}}$) \citep[e.g.][]{Shimasaku2001,Strateva2001,Kauffmann2003}. This parameter $c$ reaches a lower limit of  $\sqrt{9/5}=1.34$ for a flat light distribution.  Galaxies with more light concentrated in the inner part  usually have a larger $c$.  Early-type galaxies are usually with $ c\,\ge\,2.6$ and late-type star-forming galaxies mostly have $1.8<c<2.6$ \citep[e.g.,][]{Strateva2001,Kauffmann2003}.  The median of $c$ is 1.65 for our sample of 165 TDGs. 
Figure~\ref{fig:tdg_mc} shows stellar mass versus $c$ for  the 165 TDGs  in comparison with dwarf galaxies (DGs) adopted from \citet{Conselice2003a}.  It is clear that the TDGs are dramatically less concentrated than DGs, no cores and rather flat surface brightness distribution in the inner parts.
We note that S\'ersic models  are also often used to describe the surface brightness profile of a galaxy.  The concentration parameter $c$ is a proxy of S\'ersic index $n$. 
Fitting the PSF-corrected surface brightness profile of TDGs in our sample with a single S\'ersic function, we obtain that the majority of our TDGs are with $n < 1$, being consistent with  $c < 2.3$ for these TDGs. 
Table~\ref{tab:tdgs}, we present the stellar mass, effective radius, F814W absolute magnitude and concentration parameter of the 165 TDGs.

\begin{figure}
   \centering
	\fbox{\includegraphics[width=0.46\columnwidth]{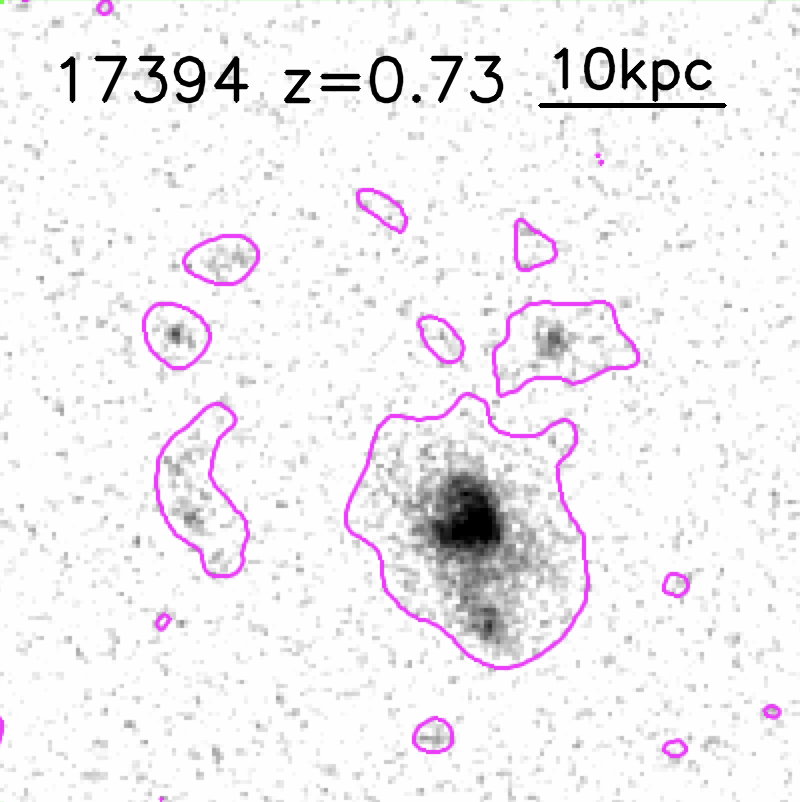}}
	\fbox{\includegraphics[width=0.46\columnwidth]{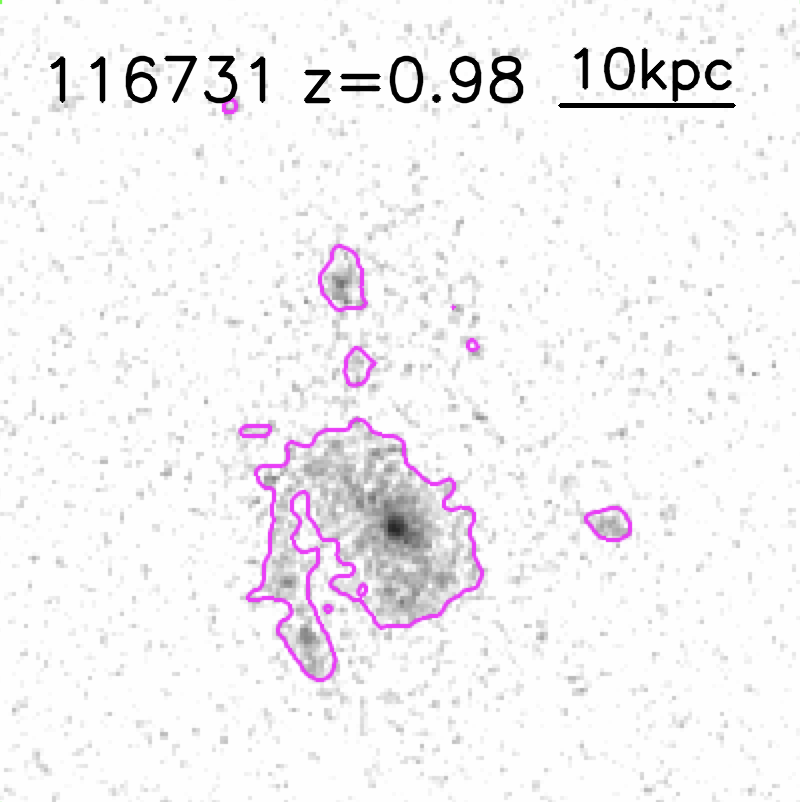}}
         \caption{{\it HST} F814W images of two merging galaxies in our sample. The contours shown with magenta lines mark the detection boundaries (1.2 times the sky background noise level). The surrounding clumps are clearly visible.}        
         \label{fig:tail_gap}
\end{figure}

We examine the mass-size relation of TDGs in comparison with different populations of dwarf galaxies. We collect samples of dwarf galaxies from the literature and show them in Figure~\ref{fig:tdg_mr}. The TDGs from our sample are comparable to those in the nearby universe given in \citet{Dabringhausen2013} and in intermediate redshifts from \citet{Wen2012}. Interestingly, the early-type satellite galaxies of the nearby galaxies from \citet{Duc2014} fall into the same region as TDGs, implying that some of the satellites might be tidally originated. 

We fit the $M_\ast - R_{\rm e}$ relation of TDGs with a power-law, giving the best fit as $\log R_{\rm e} = 0.24\ (\log (M_\ast/\rm M_\odot) -10) + 1.84$,  shown with the black-solid line in Figure~\ref{fig:tdg_mr}. We also present the $M_\ast - R_{\rm e}$ relations from \citet{Lange2016} for spiral galaxies (magenta-dashed line), elliptical galaxies (blue dot-dashed line) and blue spheroids (green-dotted line).  We can see that TDGs obey the same  $M_\ast - R_{\rm e}$ relation as elliptical galaxies and blue spheriods of similar stellar masses. In contrast, spiral galaxies are more extended than ellipticals, blue spheroids and TDGs. 
 In addition, we present ultra-diffuse galaxies  (UDGs) from \citet{Shi2017} in Figure~\ref{fig:tdg_mr} for comparison. We can see that TDGs are much smaller than UDGs of similar stellar masses. Of 165 TDGs, only 10 TDGs are with $R_{\rm e}>1.5$\,kpc. 
However, these extended TDGs have central surface brightness at least two magnitudes brighter than UDGs, as shown in Figure~\ref{fig:tdg_dis}, and therefore cannot be considered as bona-fide UDGs unless they constitute their brighter end. 
 

\begin{table}
\centering
\scriptsize
\renewcommand\tabcolsep{5.0pt} 
  \caption{The parameters of 165 tidal dwarf galaxies identified from 126 merging galaxies with long tidal tails.}
  \label{tab:tdgs}
  \centering
  \begin{tabular}{lcccccccc}
  \hline\hline
    ID & R.~A. & Dec & $z$ & $\log (M_\ast)$ & $R_{\rm e}$ & $m_{I}\,^{\rm a}$ & c$^{\rm b}$ \\
      & (J2000.0) & (J2000.0) & & $(M_\odot)$ &  (kpc) &  (mag) & \\
    \hline
TDG-218	   & 150.409721 & 1.641231 &	0.94 & 8.999	& 0.483 & 26.000 &	 1.79	\\
TDG-737	   & 150.634840 & 1.666839 &	0.85 &	 8.629	& 0.679 & 25.277 &	 1.47	\\
TDG-1692	   & 150.377754 & 1.695574 &	0.88 &	 8.685	& 1.734 & 24.813 &	 1.62	\\
TDG-4590	   & 150.625481 & 1.773197 &	0.23 &	 7.252	& 0.391 & 25.496 &	 1.68	\\
TDG-7458	   & 150.490296 & 1.849473 &	0.38 &	 8.886	& 1.102 & 23.686 &	 1.60	\\
TDG-7613	   & 150.674347 & 1.853363 &	0.51 & 8.689	& 0.491 & 24.227 &	 1.75	\\
TDG-9348-1	   & 150.485706 & 1.885998 &	0.26 &	 7.939	& 0.633 & 24.077 &	 1.78	\\
TDG-9348-2	   & 150.485753 & 1.886106 &	0.26 &	 7.679	& 0.235 & 24.728 &	 1.56	\\
TDG-13685-1  & 150.497095 & 1.961305 &	0.35 &	 7.759	& 0.245 & 25.452 &	 1.54	\\
TDG-13685-2  & 150.497189 & 1.961506 &	0.35 &	 7.796	& 0.363 & 25.360 &	 1.98	\\
... & ... & ... & ... & ... & ... & ... & ... \\
\multicolumn{8}{l}{(The machine-readable and complete version of this table is available online.)} \\
\hline\hline
\multicolumn{8}{l}{$^{\rm a}$ F814W magnitude; $^{\rm b}$concentration (i.e., $R_{\rm 50}/R_{\rm 90}$) }. \\
\end{tabular}
\end{table}

\begin{figure*}
          \centering
         \fbox{ \includegraphics[width=0.48\columnwidth]{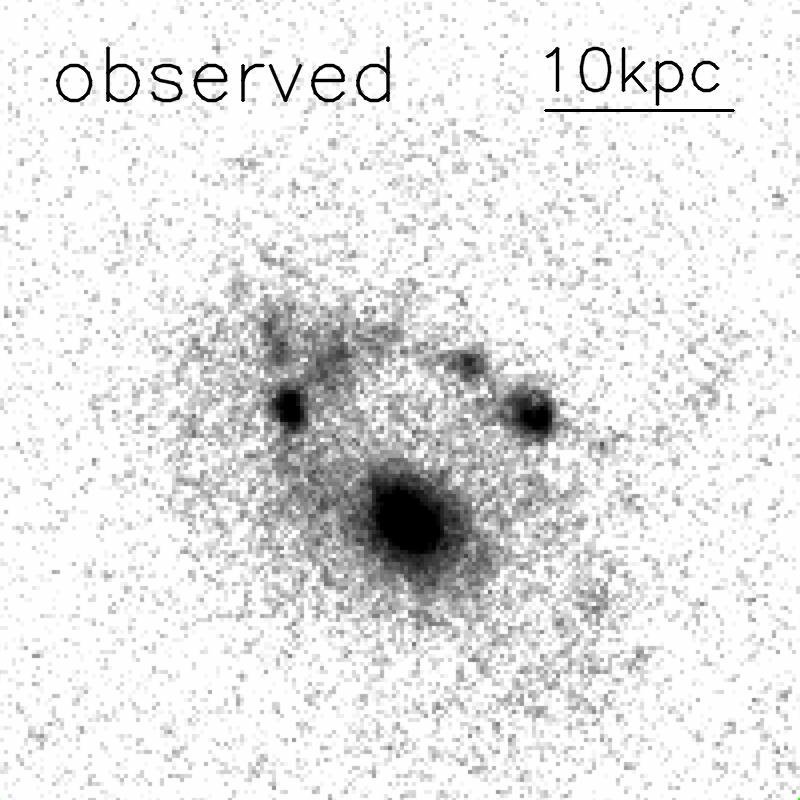}}
          \fbox{\includegraphics[width=0.48\columnwidth]{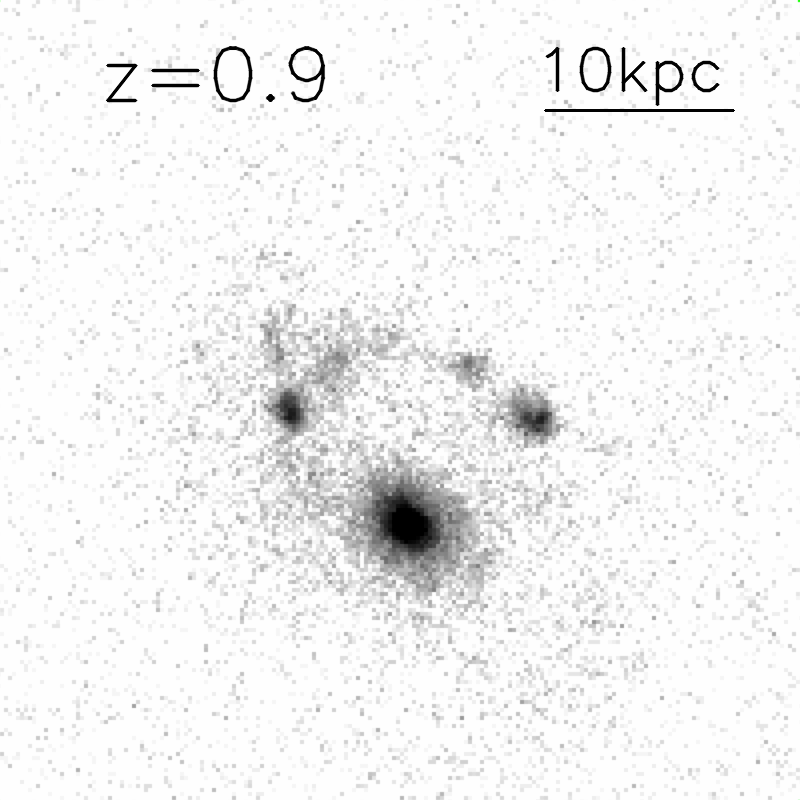}}
          \fbox{\includegraphics[width=0.48\columnwidth]{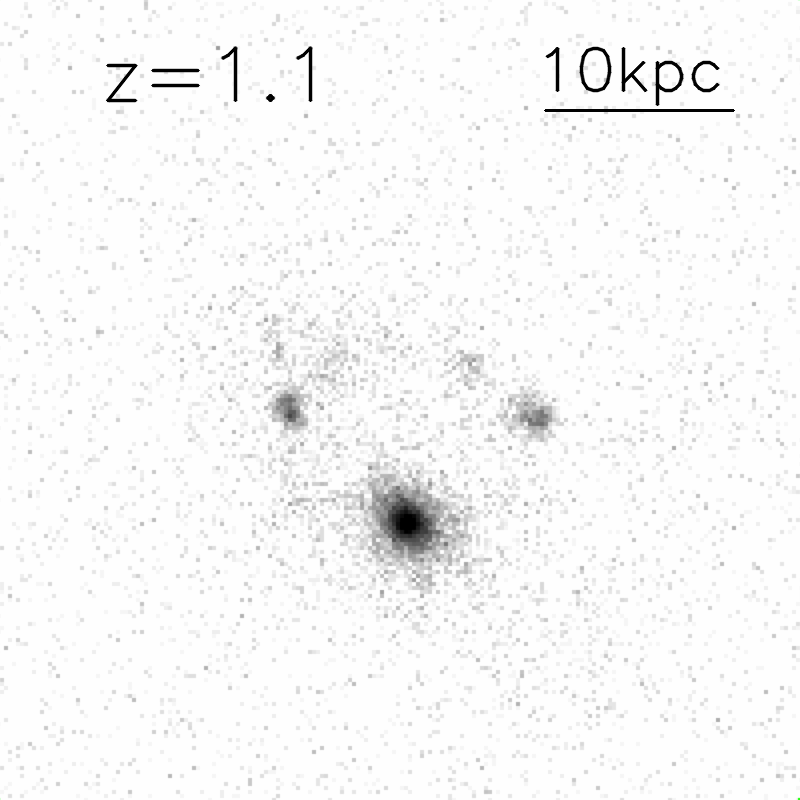}}
          \fbox{\includegraphics[width=0.48\columnwidth]{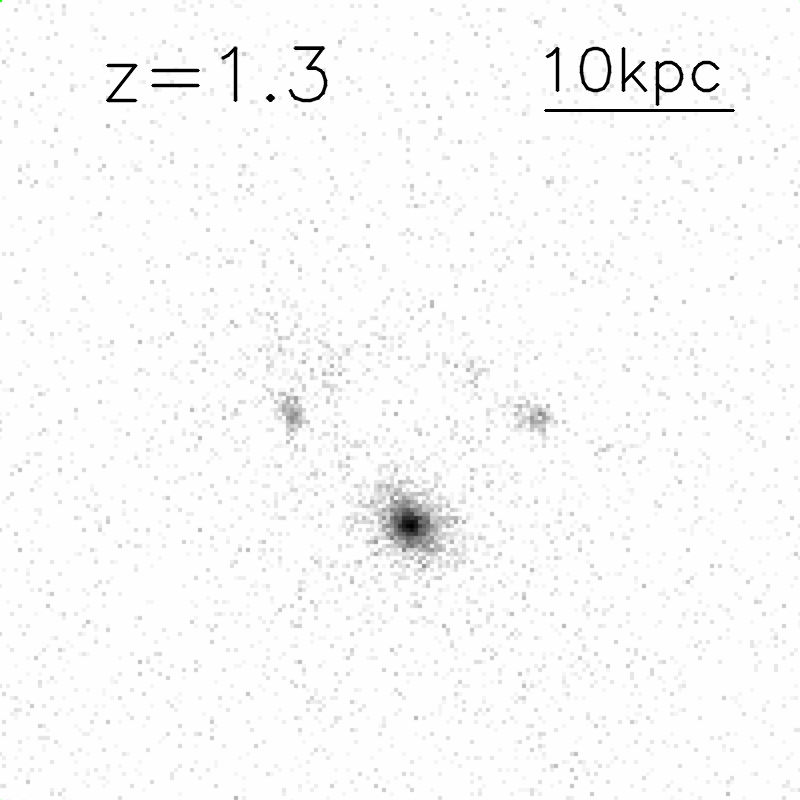}}
         \caption{{\it HST} F814W image of  a merging galaxy at $z$\,=\,0.69 (top-left), and the corresponding images at the same observational depth when redshifting it to $z$\,=\,0.90, 1.1 and 1.3 after accounting for the cosmic dimming effect. At $z$\,=\,0.69, the the surrounding tidal tail and clumps can be clearly seen. At $z>1$ it is difficult to recognize the object as a merging system.}
           \label{fig:dimming}
\end{figure*}

\subsubsection{Isolated Clumps Around the Merging Galaxies} \label{sec:333} 

Tidal tails gradually fade and eventually become undetectable after coalescence of two galaxies. However, clumps formed on the tidal tails can still be detected after the tidal tail has dissipated. In such case, one may expect to see clumps around a merging system without clearly-visible tidal tails connecting them.  Indeed, we find similar objects in our sample. Figure~\ref{fig:tail_gap} shows two merging galaxies with surrounding clumps.  The two objects exhibit clear tidal features but the diffuse component of their tidal tails are hardly detectable.  We point out that our selection for LTTGs disfavors such systems as their long tidal tails become fainter than the limiting surface brightness.

On the other hand, the cosmic dimming may lead to the same effect on the detection of the tidal tails.  We demonstrate the case by redshifting merging galaxies to higher redshifts using low-$z$ observed objects and counting for the cosmic dimming effect.  In Figure ~\ref{fig:dimming}, the left panel gives one LTTG at $z = 0.69$ from our sample;  the right three panels present the simulated images of this object redshifted to $z$\,=\,0.9, 1.1 and 1.3 with an unchanged limiting surface brightness. It can be seen that the tidal tail and the entire system gradually fade at increasing redshift and these clumps loss connections through the tail bridges at $z>1$.  The simulated object at $z>1$ can not be identified as a merging system.  
 
A clump at the tip of a tidal tail  is often barely connected. The tip clump accretes surrounding materials and quickly creates a gap to the tail.  The top two panels of Figure ~\ref{fig:tdgs} demonstrate such cases.  However, such clumps can be clearly identified.  We thus include them into our clump sample.  The tip clumps are exclusively TDGs.

\begin{figure}
         \centering
         \fbox{\includegraphics[width=0.46\columnwidth]{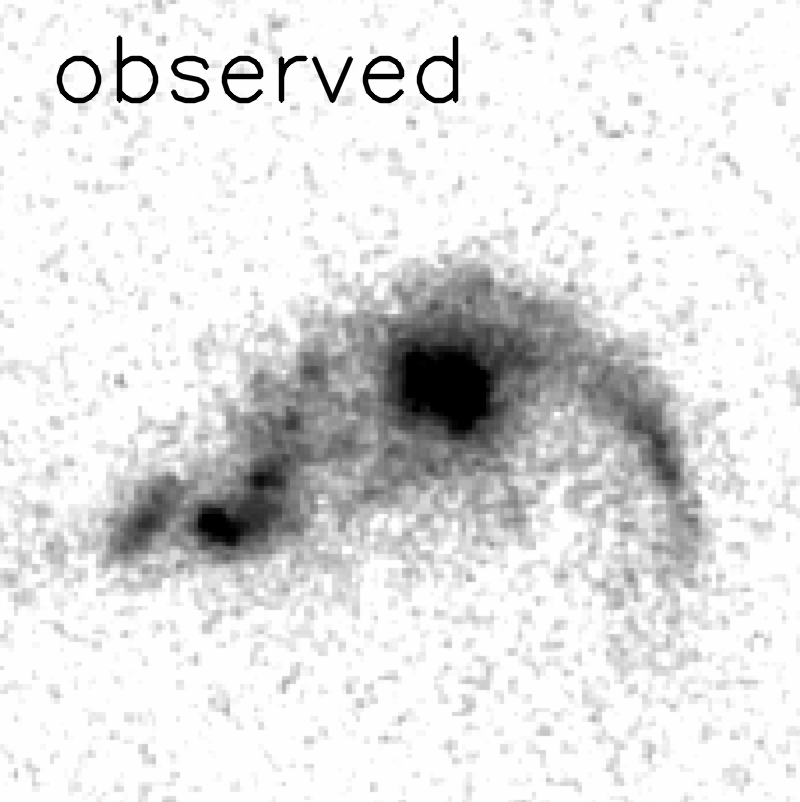}} 
         \fbox{\includegraphics[width=0.46\columnwidth]{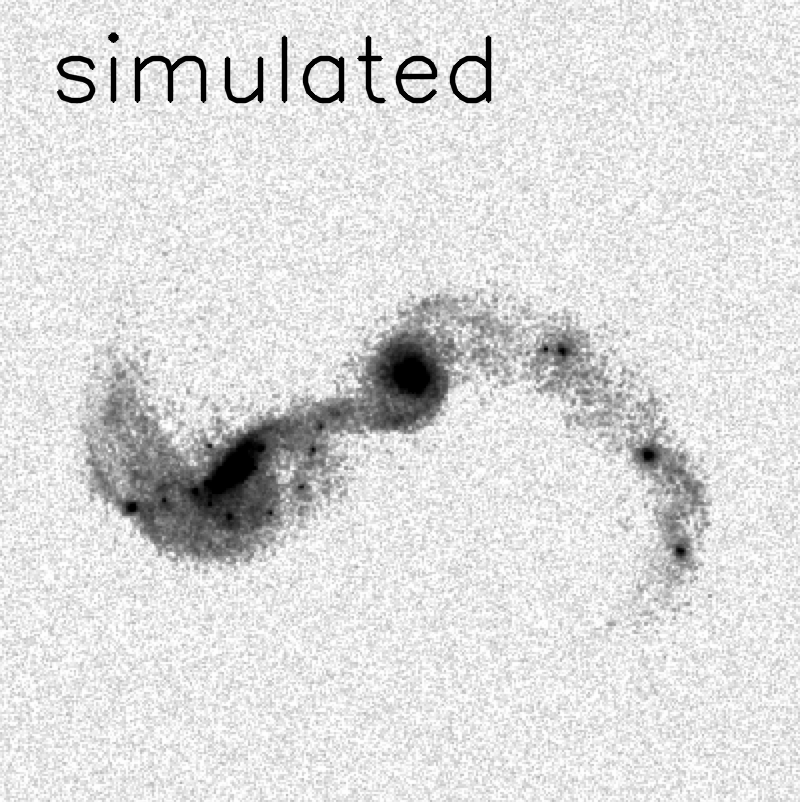}}
         \caption{Comparison of a real merger in our sample (left) with a simulated merger (right) from the GalMer database \citep{Chilingarian2010}.  The shape of the tidal tails in the two merger systems are similar.}
         \label{fig:comp}
\end{figure}


 \section{Discussion} \label{sec:discussion}

\subsection{Formation of Tidal Tails in Relation to Their Shapes}

In galaxy merging processes, the shape of tidal tails is governed by galaxy morphologies, kinematics, merger parameters, merger stage and the viewing angle \citep[e.g.][]{Toomre1972,Barnes1992}. An effective way to assess the formation of these tidal tails is to match an observed merger with simulated ones in morphology and take the merger parameters from the matched simulation to model the observed merger and formation of the associated tidal tails.  Figure~\ref{fig:comp} illustrates a real merger from our sample and a simulated merger from the GalMer database \citep{Chilingarian2010}.  The snapshot of the simulated merger is chosen to be analog to the morphology of the real merger.  This simulation models a prograde merger at the intermediate stage in which the tidal tails keep growing.  We point out that it is a big effort to pinpoint a close match of simulated mergers to each of all 461 LTTGs in our sample and figure out how these mergers were undergoing. It is worth doing but beyond the scope of this work.  Here we discuss the formation mechanisms for the three types of tidal tails in our sample. 

\begin{figure*}
          \fbox{\includegraphics[width=0.5\columnwidth]{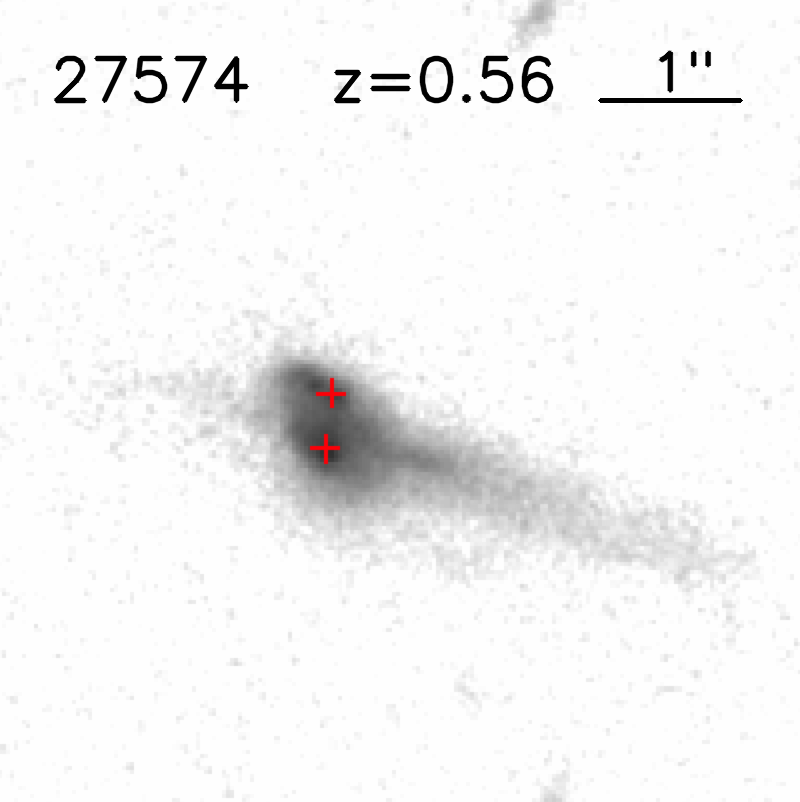}} 
          \fbox{\includegraphics[width=0.5\columnwidth]{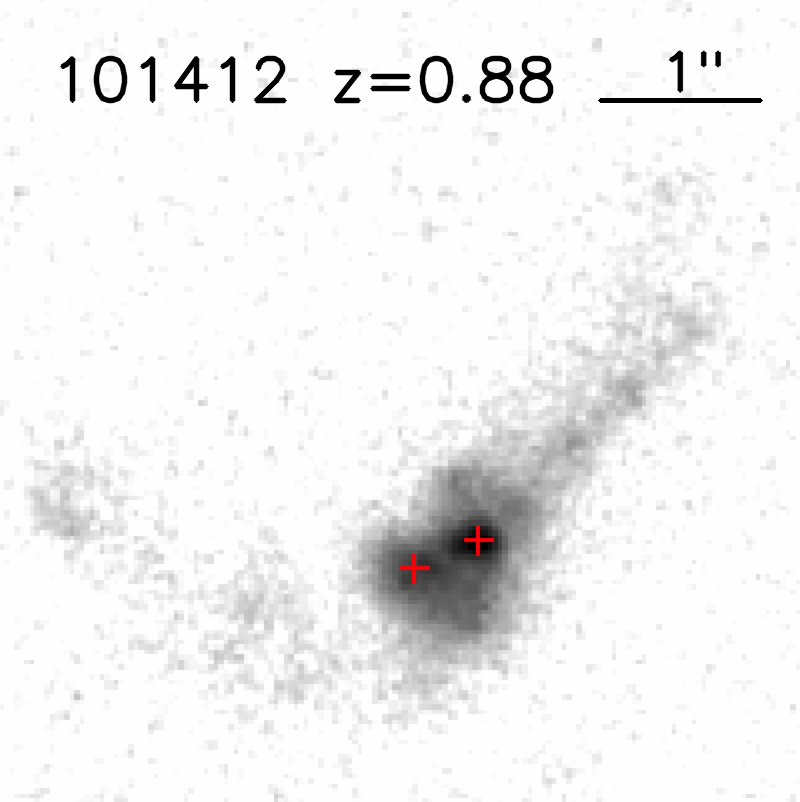}}
          \fbox{\includegraphics[width=0.5\columnwidth]{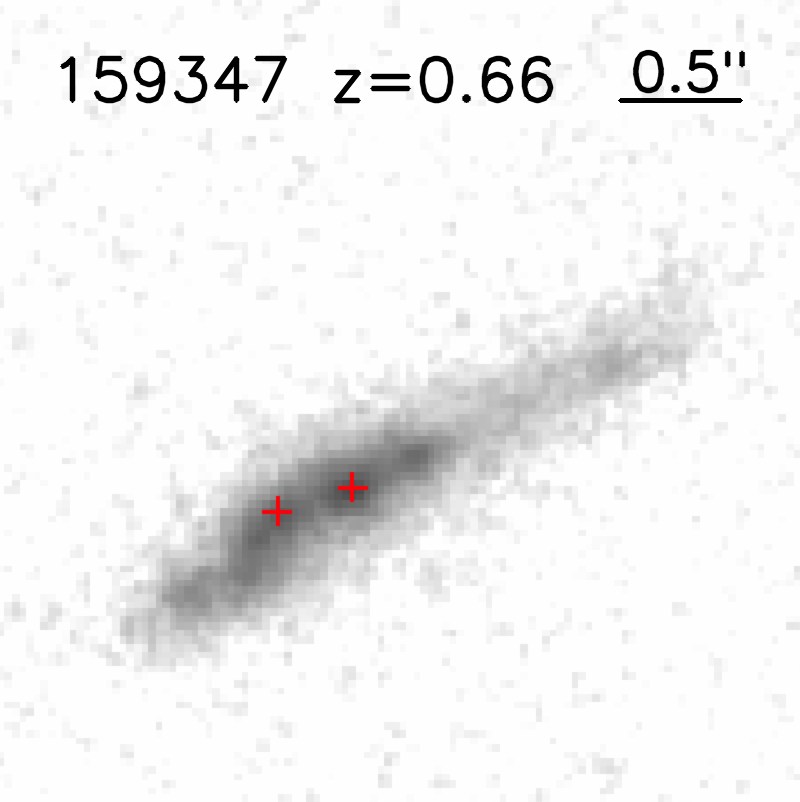}}
\caption{Three example images of straight tails. Two galaxy nuclei can be clearly identified in these systems.}
\label{fig: straight}
\end{figure*}

{\bf Straight tails} --- Our results show that the fractions of straight and curved tails are comparable in our sample.  The straight tails represent a large fraction ($\sim 40$\,per\,cent) of the overall tidal tails.  Straight tails are statistically shorter than curved tails, as shown in Figure~\ref{fig:length}. Their host galaxies are mostly  single nucleus or close pairs, are likely being in the late stages of merging processes.   In Figure~\ref{fig: straight}, we present three more examples of straight tails hosted by double-nucleus systems.   Compared with the curved tails, the straight tails tend to have a lower mean surface brightness and host less clumps as well as TDGs. 

We note that curved tidal tails mostly lie in the same plane of their rotating host galaxies. They could be viewed as straight from the edge-on angle. However, the fraction of the edge-on curved tidal tails is small if the inclination of these merging galaxies is not preferentially aligned.  And such projection effect of the tidal tails can be clarified in nearby merging galaxies \citep{Bournaud2004}.  We thus believe that the majority of straight tidal tails were formed to be straight. 

It is unclear  how to generate straight tidal tails in galaxy mergers at a high rate. From the properties of the straight tails, we suspect that they are associated with the late stages of galaxy mergers, and probably strongly stretched by violent tidal forces.  Moreover, the lack of clumps and/or relatively short lengths may hint that their parent galaxies are not sufficiently massive.  Further efforts are needed to interpret the formation of straight tidal tails.

{\bf Curved tails} --- Simulation suggest that long curved tails are often produced in the prograde mergers of disc galaxies. The length of the curved tails are significantly affected by the merger parameters such as the different inclination angle and the mass ratio. It is clear that curved tails formed after their first pericentre passage. This stage represents a relatively long period of the entire merging process \citep{Barnes1992}. 

{\bf Plume tails} --- This type of tidal tails may be associated with the retrograde mergers of disc galaxies. In retrograde merging, the materials of the tails are driven away by tidal force, and then fall back to the host galaxies and form extended plume tails. Moreover, plume tails are in place after  coalescence. Around the merger remnant, the tails and fallback materials dissipate and  form a plume appearance located at the outskirts of the host galaxies.

\subsection{Detection Completeness}

Our sample of 461 LTTGs is increasingly incomplete at $z\gtrsim 0.4$ in detecting long tidal tails down to the limiting surface brightness of 25.1\,mag\,arcsec$^{-2}$ (3\,$\sigma$ detection, \citealt{Wen2016}). Figure~\ref{fig:dimming} demonstrated the increasing detection incompleteness due to the cosmic dimming effect. \citet{Wen2016} examined the F814W surface brightness of these tidal tails and estimated the detection completeness (as shown in their figure~9). The tidal tail detection is complete above an intrinsic surface brightness of 23.1\,mag\,arcsec$^{-2}$ at $z<0.4$ after correcting for the cosmic dimming effect, and the completeness becomes  78, 72, and 62\,per\,cent  at $z\sim0.5$, 0.7 and 0.9, respectively. 

We classified the long tidal tails into three shape types. Among them,  plume tails are statistically fainter in surface brightness than straight and curved tails. The detection completeness is thus lower than for  plume tails than the other two types and our sample contains only a few plume tails at $z>0.6$. This bias lowers the fraction of the plume type in our sample.  

\begin{figure}
          \fbox{\includegraphics[width=0.46\columnwidth]{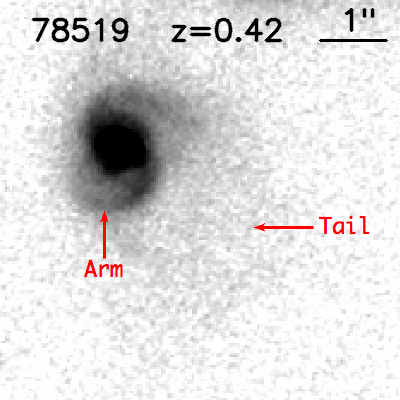}} 
          \fbox{\includegraphics[width=0.46\columnwidth]{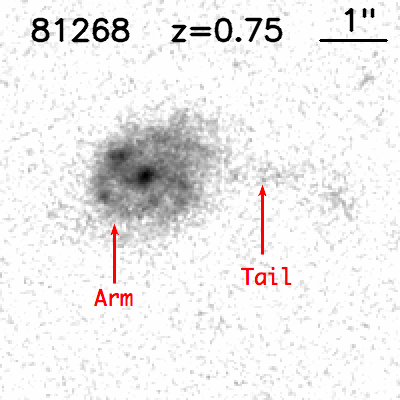}}\\
          \fbox{\includegraphics[width=0.46\columnwidth]{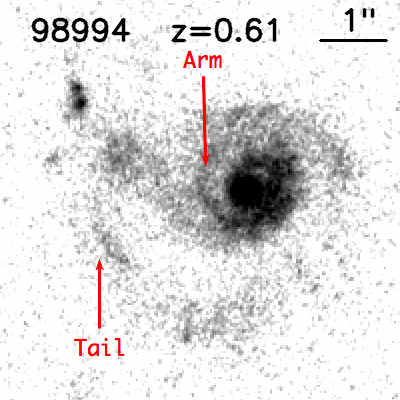}}
          \fbox{\includegraphics[width=0.46\columnwidth]{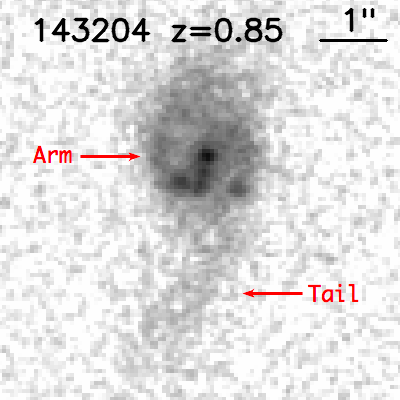}} 
         \caption{Four merging galaxies with spiral arms and tidal tails in our sample.}
         \label{fig:spiral}
\end{figure}

\subsection{Formation of TDGs}

We identified 165 TDGs in 126 LTTGs. This is so far the largest sample of TDGs at $0.2<z<1$.  The properties of TDGs in our sample, as shown in Figure~\ref{fig:tdg_dis}, are generally consistent with those reported in previous works for individual cases in the nearby universe or small samples of TDGs.  Based on the to date largest TDG sample, we estimated a production rate of TDGs to be 0.36 per major merger. 
In order to estimate the contribution of TDGs to the dwarf population in the local universe, we selected a subsample of TDGs with  $8.5\leq\log (M_\ast/ \rm M_\odot)\leq9.5$ and calculate the production rate of TDGs in this mass range. The production rate of TDGs in this mass range is $\sim$\,0.17 TDGs per long-tidal tail merger system.
Combining galaxy mass functions at $z<4$ from \citet{Laigle2016} with the LTTG merger fraction from \citet{Wen2016}, we obtain a TDG density function  in the mass range of $\log (M_\ast/\rm M_\odot)\geq9.5$. Adopting the TDG survival rates from \citet{Bournaud2006} , we calculate the accumulated number density of survived TDGs  to be $1.2\times10^{-4}\,{\rm Mpc^{-3}}$ over the past $1-12$\,Gyr  and $1.7\times10^{-4}\,{\rm Mpc^{-3}}$ over $0-1$\,Gyr lookback time.   

Considering that some TDGs in $0-1$\,Gyr do not have enough time to escape from tidal tails and evolve into isolated dwarf galaxies, we find that the number density of dwarf galaxies evolved from TDGs since 12\,Gyr ago is about $2\times10^{-4}\,{\rm Mpc^{-3}}$.
From the local galaxy mass function \citep{Bell2003}, we  obtain that dwarf galaxies with $8.5\leq\log (M_\ast/\rm M_\odot)\leq9.5$ have a number density of about $7\times10^{-3}\,{\rm Mpc^{-3}}$  in the local universe. This means that about $\sim3$ per\,cent of dwarf galaxies are contributed by TDGs.  Considering that half of the merging galaxies have long tidal tails and not all of the merging systems can produce TDGs \citep{Bournaud2006,Wen2016}, we estimate that $\sim5$ per\,cent of dwarf galaxies are of tidal origin. This is consistent with the result given in \citet{Wen2012}. Such a low production rate indicates that the tidally-originated dwarf galaxies do not represent a significant fraction of the dwarf population in the same stellar mass regime.

We denote that most of the TDGs are associated with curved (85/165)  and straight (55/165) tails. Interestingly, about half of TDGs are located at the tip of their host tails. We compare the tidal tails hosting TDGs with those containing no TDGs, finding no systematic difference in their properties. We suspect that gas fraction might be important for triggering the formation of TDGs \citep[e.g.,][]{Duc2004}.  Our current investigation is unable to tell the reasons why some tidal tails can form TDGs at the tip while the others do not.  Simulations of galaxy mergers can help to understand the formation mechanisms of TDGs. 

TDGs are believed to contain little dark matter because of their tidal origin \citep{Bournaud2006}.  Our TDGs have spherical-like morphologies with stellar mass mostly in the range of $\log (M_\ast/\rm M_\odot)\sim 7.5-9.5$ and half-light radius around $R_{\rm e}\sim 0.8$\,kpc.  Our results show that TDGs follow the mass-size  relation of ellipticals and spheroids in the same stellar mass range (Figure~\ref{fig:tdg_mr}). This coincidence suggests that at the low-mass end the central part of ellipticals and spheroids are likely dominated by the baryonic matter. In addition, our results reveal that the average surface brightness profile of TDGs is globally flatter than that of disc galaxies. We suspect that TDGs in our sample are still in the formation process and not dynamically relaxed yet, making them even more extended than disc galaxies. It is worth noting that  the properties of TDGs, i.e., relatively small size and low S\'ersic index,  are similar to those of high-$z$ star-forming galaxies.  TDGs could be thus used to explore the physical processes of rapid star formation that more often occur in high-$z$ galaxies.  

Recently,  some UDGs were claimed to be dominated by baryonic matter and possibly tidal originated.  
We examine our TDGs  to see if some of them could be the UDG candidates. We note that UDGs are defined to have $R_{\rm e}>$\,1.5\,kpc and the central surface brightness of 24 $-$ 26\,mag\,arcsec$^{-2}$ \citep[e.g.][]{Koda2015,Shi2017}.   In our sample of 165 TDGs,  there are only  10 with $R_{\rm e}>1.5$\,kpc. Their central surface brightness are, however, at least two magnitudes brighter than that of UDGs.  Indeed,  the limiting surface brightness of the COSMOS {\it HST} imaging data does not allow us to detect UDGs associated with our sample LTTGs.  On the other hand, all of the TDGs on tidal tails are likely forming new stars through attracting surrounding gas and thus show a high brightness temporarily. We cannot deny the possibility that at least some of the 10 extended UDGs would fade dramatically and satisfy the criteria for being UDGs. This is supported by the finding that some old TDGs exhibit properties similar to those of UDGs \citep[e.g.,][]{Duc2014}.

\subsection{Spiral Arms and Curved Tails}

Spiral arms in (barred) disc galaxies, in particular the ``loose'' ones, appear like curved tails and might contaminate our sample. We argue that the contamination is marginal and our results are not affected. 
Our sample galaxies are selected to have disturbed morphologies with strong tidal features in the outskirts.  As shown in Figure~\ref{fig:tailtype}, \ref{fig:clump} and \ref{fig:tdgs}, the long tidal tails in these galaxies are asymmetric and extended. We also check all 461 LTTGs and pick up these with arm structures.  Figure~\ref{fig:spiral} shows the {\it HST} images of four sample galaxies with the least disturbed spiral arms. Both spiral arms and tidal tails can be seen in these galaxies. The spiral arms are symmetric and closely linked with the core of the host galaxies. In contrast, the tidal tails are diffuse and more extended.  In some merging pairs of galaxies with "loose'' spiral arms, tidal forces may drive the arms open and turn them into curved tails. This mechanism does not require strong tidal forces to break the original disc structures, likely responsible for part of the curved tails in galaxy pairs with nuclear separation distance $d>20$\,kpc. 

\section{Conclusions}

Using {\it HST}/ACS F814W imaging data in conjunction with publicly-available multi-wavelength data  in the COSMOS field,  we analyse the morphological properties for a sample of 461 merging galaxies with long tidal tails. We summarize our results as follows: 

\begin{itemize}
\item   Of 461 LTTGs,  91 (21\,per\,cent) are single-nucleus merging galaxies, 90 (20\,per\,cent) are partially-coalesced with nuclear separation distance $d<20$\,kpc, 128 (28\,per\,cent) are close pairs with $d<20$\,kpc, and 145 (31\,per\,cent) are galaxy pairs with  $d>20$\,kpc. Comparison of our sample galaxies with  simulations suggests that long tidal tails are mostly generated in prograde galaxy mergers.  

\item The long tidal tails can be classified mainly into three shape types:   straight,  curved and  plume.  Our sample consists of 192 (41\,per\,cent) LTTGs with straight tails,  213 (47\,per\,cent) with curved tails, and  56 (12\,per\,cent) with plume tails. We find that  the curved tails viewed edge-on may be responsible for only a small fraction of the straight tails.  It remains to be understood how galaxies merge to generate straight tails at such a high rate.  We argue that the shape of tidal tails is generally coupled with the stage of merging processes although it also relies on the merger parameters. 

\item  We identify 177 clumps from all  long tidal tails of our 461 merging galaxies.  Of them, 165 are classified as tidal dwarf galaxies residing in 126 host merging galaxies, yielding a production rate of 0.36. We estimate that $\sim$\,5\,per\,cent of local dwarf galaxies  are tidally originated.  Our results show that most of the TDGs are associated with curved (85/165)  and straight (55/165) tails.  More than half of TDGs are located at the tip of their host tails. 

\item We find that TDGs follow the mass-size relation of local elliptical galaxies over $7.5\leq\log(M_\ast/$M$_\odot)\leq9.5$, while their  surface brightness profiles appear statistically flatter than an exponential disc.  These indicate that TDGs are still in the formation process, make them analogous objects for studying rapid star formation that commonly occur in high-$z$ star-forming galaxies. 

\end{itemize}

\section*{Data availability}
The data underlying this article are available in the article and in its online supplementary material.
\section*{Acknowledgements}

We thank the  anonymous referee for her/his useful comments and suggestions that improved this manuscript.  
This work is supported by the National Key Research and Development Program of China (2017YFA0402703), and the National Science Foundation of China (11773076, 11703092, 12073078), and the Chinese Academy of Sciences (CAS) through a China-Chile Joint Research Fund (CCJRF \#1809) administered by the CAS South America Centre for Astronomy (CASSACA).  DVG gratefully acknowledges the support from the CAS President's International Fellowship Initiative (PIFI) visiting scientists program (2019VMA0003).

\bsp	
\label{lastpage}

\end{document}